\documentclass[twoside,11pt,nohyperref]{article}
\usepackage{subcaption}
\usepackage{jmlr2e}
\usepackage[USenglish]{babel}
\usepackage[utf8]{inputenc}
\usepackage[T1]{fontenc}
\usepackage{amsmath}
\usepackage{extdash}
\usepackage{bm}
\usepackage{pgfplots}
\usepackage[per-mode=symbol]{siunitx}
\usepackage{placeins}
\usepackage{lastpage}

\jmlrheading{20}{2019}{1-\pageref{LastPage}}{7/18; Revised
	1/19}{2/19}{18-476}{Maximilian H\"uttenrauch and Adrian \v{S}o\v{s}i\'{c} and Gerhard Neumann}
\ShortHeadings{Deep Reinforcement Learning for Swarm Systems}{H\"uttenrauch, \v{S}o\v{s}i\'{c} and Neumann}

\firstpageno{1}

\newcommand{\given}{\,|\,}
\newcommand{\einschub}[1]{\Emdash#1\Emdash}
\newcommand{\textslash}{\,/\,}
\begin{document}
	
	\title{Deep Reinforcement Learning for Swarm Systems
	}
	
	\author{\name Maximilian H\"uttenrauch \email mhuettenrauch@lincoln.ac.uk \\
		\addr L-CAS\\
		University of Lincoln\\
		LN6 7TS Lincoln, UK
		\AND
		\name Adrian \v{S}o\v{s}i\'{c} \email adrian.sosic@bcs.tu-darmstadt.de \\
		\addr Bioinspired Communication Systems\\
		Technische Universit\"at Darmstadt\\
		64283 Darmstadt, Germany
		\AND
		\name Gerhard Neumann \email gneumann@lincoln.ac.uk \\
		\addr L-CAS\\
		University of Lincoln\\
		LN6 7TS Lincoln, UK
		}
	
	\editor{George Konidaris}
	
	\maketitle

\begin{abstract}%
Recently, deep reinforcement learning (RL) methods have been applied successfully to multi-agent scenarios. 
Typically, the observation vector for decentralized decision making is represented by a concatenation of the (local) information an agent gathers about other agents.
However, concatenation scales poorly to swarm systems with a large number of homogeneous agents as it does not exploit the fundamental properties inherent to these systems: (i) the agents in the swarm are interchangeable and (ii) the exact number of agents in the swarm is irrelevant.  
Therefore, we propose a new state representation for deep multi-agent RL based on mean embeddings of distributions, where we treat the agents as samples and use the empirical mean embedding as input for a decentralized policy.
We define different feature spaces of the mean embedding using histograms, radial basis functions and neural networks trained end-to-end.
We evaluate the representation on two well-known problems from the swarm literature\einschub{rendezvous and pursuit evasion}in a globally and locally observable setup.
For the local setup we furthermore introduce simple communication protocols.
Of all approaches, the mean embedding representation using neural network features enables the richest information exchange between neighboring agents, facilitating the development of complex collective strategies.
\end{abstract}


\section{Introduction}
\label{intro}
In swarm systems, many identical agents interact with each other to achieve a common goal.
Typically, each agent in a swarm has limited capabilities in terms of sensing and manipulation so that the considered tasks need to be solved collectively by multiple agents.

A promising application where intelligent swarm systems take a prominent role is swarm robotics \citep{bayindir2016review}.
Robot swarms are formed by a large number of cheap and easy to manufacture robots that can be useful in a variety of situations and tasks, such as search and rescue missions or exploration scenarios.
A swarm of robots is inherently redundant towards loss of individual robots since usually none of the robots plays a specific role in the execution of the task. Because of this property, swarm-based missions are often favorable over single-robot missions (or, let alone, human missions) in hazardous environments.
Behavior of natural swarms, such as foraging, formation control, collective manipulation, or the localization of a common `food' source can be adapted to aid in these missions \citep{bayindir2016review}.
Another field of application is routing in wireless sensor networks \citep{saleem2011swarm} since each sensor in the network can be treated  as an agent in a swarm.

A common method to obtain control strategies for swarm systems is to apply opti\-mization-based approaches using a model of the agents or a graph abstraction of the swarm \citep{lin2004local, jadbabaie2003coordination}.
Optimization-based approaches allow to compute optimal control policies for tasks that can be well modeled, such as rendezvous or consensus problems \citep{lin2007multi} and formation control \citep{ranjbar2012novel}, or to learn pursuit strategies to capture an evader \citep{zhou2016cooperative}.
Yet, these approaches typically use simplified models of the agents\textslash the task and often rely on unrealistic assumptions, such as operating in a connected graph \citep{dimarogonas2007rendezvous} or having full observability of the system state \citep{zhou2016cooperative}.
Rule-based approaches use heuristics inspired by natural swarm systems, such as ants or bees \citep{handl2007ant}.
Yet, while the resulting heuristics are often simple and can lead to complex swarm behavior, the obtained rules are difficult to adapt, even if the underlying task changes only slightly.

Recently, deep reinforcement learning (RL) strategies have become popular to solve multi-agent coordination problems.
In RL, tasks are specified indirectly through a cost function, which is typically easier than defining a model of the task directly or a finding a heuristic for the controller. 
Having defined a cost function, the RL algorithm aims to find a policy that minimizes the expected cost.   
Applying deep reinforcement learning within the swarm setting, however, is challenging due to the large number of agents that need to be considered.
Compared to single-agent learning, where the agent is confronted only with observations about its own state, each agent in a swarm can make observations of several other agents populating the environment and thus needs to process an entire set of information that is potentially  varying in size.
Accordingly, two main challenges can be identified in the swarm setting:
\begin{enumerate}
	\item High state and observation dimensionality, caused by large system sizes.
	\item Changing size of the available information set, either due to addition or removal of agents, or because the number of observed neighbors changes over time.
\end{enumerate}
Most current multi-agent deep reinforcement learning methods either concatenate the information received from different agents \citep{lowe2017multi} or encode it in a multi-channel image, where the image channels contain different features based on a local view of an agent \citep{sunehag2017value, zheng2017magent}.
However, both types of methods bare major drawbacks.
Since neural network policies assume a fixed input dimensionality, a concatenation of observations is unsuitable in the case changing agent numbers.
Furthermore, a concatenation disregards the inherent permutation invariance of identical agents in a swarm system and scales poorly to large system sizes.
Top-down image based representations alleviate the issue of permutation invariance, however, the information obtained from neighboring agents is of mostly spatial nature.
While additional information can be captured by adding more image channels, the dimensionality of the representation increases linearly with each feature.
Furthermore, the discretization into pixels has limited accuracy due to quantization errors.

\vspace{\baselineskip}
In this paper, we exploit the homogeneity of swarm systems and treat the state information perceived from neighboring agents as samples of a random variable. Based on this model, we then use mean feature embeddings (MFE) \citep{smola2007hilbert} to encode the current distribution of the agents.
Each agent gets a local view of this distribution, where the information obtained from the neighbors is encoded in the mean embedding.
Due to the sample-based view of the collected state information, we achieve a permutation invariant representation that is furthermore invariant to the number of agents in the swarm\textslash the number of perceived neighbors.

Mean feature embeddings have so far been used mainly for kernel-based feature representations \citep{gretton2012kernel}, but they can be also applied to histograms or radial basis function (RBF) networks.
The resulting models are closely related to the ``invariant model'' formulated by \citet{zaheer2017deep}. However, compared to the summation approach described in their paper, the averaging of feature activations proposed in our approach yields the desired invariance with respect to the observed agent number mentioned above.
To the best of our knowledge, we are the first to use mean embeddings inside a deep reinforcement learning framework for swarm systems where both the feature space of the mean embedding as well as the policy are learned end-to-end.

We test our state representation on various rendezvous and pursuit evasion problems using Trust Region Policy Optimization (TRPO) \citep{schulman2015trust} as the underlying deep RL algorithm.
In the rendezvous problem, the agents need to find a collective strategy that allows them to meet at some arbitrary location.
In the pursuit evasion domain, a group of agents collectively tries to capture one or multiple evaders.

Policies are learned in a \emph{centralized-learning\textslash decentralized-execution fashion} fashion, meaning that during learning data from all agents is collected centrally and used to optimize the parameters as if there was only one agent. Nonetheless,  each agent only has access to its own perception of the global system state to generate actions from the policy function.
We compare our representation to several deep RL baselines as well as to optimization-based solutions, if available.
Herein, we perform our experiments both in settings with global observability (i.e., all agents are neighbors) and in settings with local observability (i.e., agents are only locally connected).
In the latter setting, we also evaluate different communication protocols \citep{huettenrauch2018local} that allow the agents to transmit additional information about their local graph structure.
For example, an agent might transmit the number of neighbors within its current neighborhood.
Previously, such additional information could not be encoded efficiently due to the poor scalability of the histogram-based approaches.

Our results show that agents using our representation can learn faster and obtain policies of higher quality, suggesting that the representation as mean embedding is an efficient encoding of the global state configuration for swarm-based systems.
Moreover, mean embeddings are simple to implement inside existing neural network architectures and can be applied to any deep RL algorithm, which makes the approach applicable in a wide variety of scenarios.
The source code to reproduce our results can be found online.\footnote{\url{https://github.com/LCAS/deep_rl_for_swarms}}

\section{Related Work}
\label{sec:rel_work}
The main contribution of this work lies in the development of a compact representation of state information in swarm systems, which can easily be used within deep multi-agent reinforcement learning (MARL) settings that contain homogeneous agent groups.
In fact, our work is mostly orthogonal to other research in the field of MARL and
the presented ideas can be incorporated into most existing approaches. 
To provide an overview, we begin with a brief survey of algorithms used in (deep) MARL, we revisit the basics of mean embedding theory, and we summarize some classic approaches to swarm control for the rendezvous and pursuit evasion task.

\subsection{Deep RL}
Recently, there has been increasing interest in deep reinforcement learning for swarms and multi-agent systems in general.
For example, \citet{zheng2017magent} provide a many-agent reinforcement learning platform based on a multi-channel image state representation, which uses Deep Q-Networks (DQN) \citep{mnih2015human} to learn decentralized control strategies in large grid worlds with discrete actions.
\citet{gupta2017cooperative} show a comparison of centralized, concurrent and parameter sharing approaches to cooperative deep MARL, using TRPO \citep{schulman2015trust}, 
DDPG \citep{lillicrap2015continuous} and DQN.
They evaluate each method on three tasks, one of which is a pursuit task in a grid world using bitmap-like images as state representation.
A variant of DDPG for multiple agents in Markov games using a centralized action-value function is provided by \cite{lowe2017multi}.
The authors evaluate the method on tasks like cooperative communication, navigation and others.
The downside of a centralized action-value function is that the input space grows linearly with the number of agents, and hence, their approach scales poorly to large system sizes. 
A more scalable approach is presented by \cite{yang2018mean}.
Employing mean field theory, the interactions within the population of agents are approximated by the interaction of a single agent with the average 
effect from the overall population, which has the effect that the action-value function input space stays constant.
Experiments are conducted on a Gaussian squeeze problem, an Ising model, and a mixed cooperative-competitive battle game. Yet, the paper does not address
the state representation problem for swarm systems.

\citet{omidshafiei2017deep} investigate hysteretic Q-learning \citep{matignon2007hysteretic} and distillation \citep{rusu2015policy}. 
They use deep recurrent Q-networks \citep{hausknecht2015deep} to solve single and multi-task Dec-POMDP problems.
Following this work, \citet{palmer2017lenient} add leniency \citep{panait2006lenient} to the hysteretic approach to prevent ``relative overgeneralization'' of agents.
The approach is evaluated on a coordinated multi-agent object transportation problem in a grid world with stochastic rewards.

\citet{sunehag2017value} tackle the ``lazy agent'' problem in cooperative MARL with a single team reward by training each agent with a learned additive decomposition of a value function based on the team reward.
Experiments show an increase in performance on cooperative two-player games in a grid world.
\citet{rashid2018qmix} further develop the idea with the insight that a full factorization of the value function is not necessary.
Instead, they introduce a monotonicity constraint on the relationship between the global value function and each local value function.
Results are presented on the StarCraft micro management domain.

Finally, \citet{grover2018learning} show a framework to model agent behavior as a representation learning problem.
They learn an encoder-decoder embedding of agent policies via imitation learning based on interactions and evaluate it on a cooperative particle world \citep{mordatch2018emergence} and a competitive two-agent robo sumo environment \citep{al-shedivat2018continuous}.
The design of the policy function in the approach of \citet{mordatch2018emergence} is similar to ours but the model uses a softmax pooling layer. However, instead of applying (model-free) reinforcement learning to optimize the parameters of the policy function, they build an end-to-end differentiable model of all agent and environment state dynamics and calculate the gradient of the return with respect to the parameters via backpropagation.

An application related to our approach can be found in the work by \cite{gebhardt2018learning}, where the authors use mean embeddings to learn a centralized controller for object manipulation with robot swarms.
Here, the key idea is to directly embed the swarm configuration into a reproducing kernel Hilbert space, whereas our approach is based on embedding the agent's local view.
Furthermore, using kernel-based feature spaces for the mean embedding scales poorly in the number of samples and in the dimensionality of the embedded information.

\subsection{Optimization-Based Approaches for Swarm Systems}
To provide a concise summary of the most relevant related work, we concentrate on opti\-mization-based approaches that derive decentralized control strategies for the rendezvous and pursuit evasion problem considered in this paper.
\citet{ji2007distributed} derive a control mechanism preserving the connectedness of a group of agents with limited communication abilities for the rendezvous and formation control problem.
The method focuses on high-level control with single integrator linear state manipulation and provides no rules for agents that are not part of the agent graph.
Similarly, \citet{de2006decentralized} present a decentralized algorithm to maximize the connectivity (characterized by an exponential model) of a multi-agent system.
The algorithm is based on the minimization of the second smallest eigenvalue of the Laplacian of the proximity graph.
An approach providing a decentralized control strategy for the rendezvous problem for nonholonomic agents can be found in the work by \cite{dimarogonas2007rendezvous}.
Using tools from nonsmooth Lyapunov theory and graph theory, the stability of the overall system is examined.
A control strategy for the pursuit evasion problem with multiple pursuers and single evader that we investigate in more detail later in this paper was proposed \cite{zhou2016cooperative}.
The authors derive decentralized control policies for the pursuers and the evader based on the minimization of Voronoi partitions.
Again, the control mechanism is for high-level linear state manipulation.
Furthermore, the method assumes visibility of the evader at all times.
A survey on pursuit evasion in mobile robotics in general is provided by \cite{chung2011search}.\\

\subsection{Analytic Approaches}
Another line of work concerned with the curse of dimensionality can be found in the area of multi-player reach-avoid games. \citet{chen2017multiplayer}, for example, look at pairwise interactions between agents. This way, they are able to use the Hamilton-Jacobian-Isaacs approach to solve a partial differential equation in the joint state space of the players. Similar work can be found in \citep{chen2014multiplayer, chen2014path, zhou2012general}.

\section{Background}
\label{sec:background}
In this section, we give a short overview of Trust Region Policy Optimization and mean embeddings of distributions.

\subsection{Trust Region Policy Optimization}
Trust Region Policy Optimization is an algorithm to optimize control policies in single-agent reinforcement learning problems \citep{schulman2015trust}.
These problems are formulated as Markov decision processes (MDPs), which can be compactly written as a tuple $\langle \mathcal{S}, \mathcal{A}, P, R \rangle$.
In an MDP, an agent chooses an action $a \in {\mathcal{A}}$ according to some policy~$\pi(a \given s)$ based on its current state $s \in \mathcal{S}$ and progresses to state $s'\in\mathcal{S}$ according to the transition dynamics $P$, i.e., $s'\sim P(s' \given s, a)$.
After each step, the agent receives a reward $r = R(s, a)$, provided by the reward function $R$, which judges the quality of its decision.
The goal of the agent is to find a policy that maximizes the cumulative reward achieved over a certain period of time.

In TRPO, the policy is parametrized by a parameter vector~$\theta$ containing the weights and biases of a neural network.
In the following, we denote this parametrized policy as $\pi_\theta$.
The reinforcement learning objective is expressed as finding a new policy that maximizes the expected advantage function of the current policy $\pi_{\textrm{old}}$, i.e., ${J^{\text{TRPO}} = \mathbb{E}\left[\frac{\pi_{\theta}}{\pi_{\theta_{\text{old}}}} A^{\pi_{\textrm{old}}}(s, a)\right]},$ where 
$A^{\pi_{\textrm{old}}}(s,a) = Q^{\pi_{\textrm{old}}}(s,a)-V^{\pi_\textrm{old}}(s)$.
Herein, the state-action value function $Q^{\pi_{\textrm{old}}}(s,a)$ is typically estimated via trajectory rollouts, while for the value function $V^{\pi_{\textrm{old}}}(s)$ linear or neural network baselines are used that are fitted to the Monte-Carlo returns, resulting in an estimate $\hat{A}(s, a)$ for the advantage function.
The objective is to be maximized subject to a fixed constraint on the Kullback-Leibler (KL) divergence of the policy before and after the parameter update, which ensures that the updates to the policy parameters~$\theta$ are bounded, in order to avoid divergence of the learning process.
The overall optimization problem is summarized as
\begin{equation*}
\begin{aligned}
& \underset{\theta}{\text{maximize}}
& & \mathbb{E}\left[\frac{\pi_{\theta}}{\pi_{\theta_{\text{old}}}}\hat{A}(s, a)\right] \\
& \text{subject to}
& & \mathbb{E}[D_{\text{KL}}(\pi_{\theta_{\text{old}}}|| \pi_{\theta})] \leq \delta.
\end{aligned}
\end{equation*}
The problem is approximately solved using conjugate gradient optimization, after linearizing the objective and quadratizing the constraint.

\subsection{Mean Embeddings}
\label{sec:mean_embs}
Our work is inspired by the idea of embedding distributions into reproducing kernel Hilbert spaces \citep{smola2007hilbert} from where we borrow the concept of mean embeddings.
A probability distribution $P(X)$ can be represented as an element in a reproducing kernel Hilbert space by its expected feature map (i.e., the mean embedding),
\begin{align*}
\mu_X = \mathbb{E}_X[\phi(X)],
\end{align*}
where $\phi(x)$ is a (possibly infinite dimensional) feature mapping.
Given a set of observations $\{x_1, \dots, x_m\}$, drawn i.i.d.\ from $P(X)$, the empirical estimate of the expected feature map is given by
\begin{align*}
\hat{\mu}_X = \frac{1}{m} \sum_{i=1}^{m}\phi(x_i).
\end{align*}
Using characteristic kernel functions $k(x, x') = \langle \phi(x),~\phi(x') \rangle$, such as Gaussian RBF or Laplace kernels, mean embeddings can be used, for example, in two-sample tests \citep{gretton2012kernel} and independence tests \citep{gretton2008kernel}.
A characteristic kernel is required to uniquely identify a distribution based on its mean embedding.
However, this assumption can be relaxed to using finite feature spaces if the objective is merely to extract relevant information from a distribution such as, in our case, the information needed for the policy of the agents.

\section{Deep Reinforcement Learning for Swarms}
\label{sec:swarmdrl}
The reinforcement learning algorithm presented in the last section has been originally designed for single-agent learning.
In order to apply this algorithm to the swarm setup, we switch to a different problem domain and show the implications on the learning algorithm.
Policies in this context are then optimized in a centralized--learning\textslash decentralized--execution fashion.

\subsection{Problem Domain}
\label{sec:prob_state}
The problem domain for our swarm system is best described as a swarm MDP environment \citep{sosic2017inverse}.
The swarm MDP can be regarded as a special case of a decentralized partially observable Markov decision process (Dec-POMDP) \citep{bernstein2002complexity} and is constructed in two steps. 
First, an agent prototype is defined as a tuple $\mathbb{A} = \langle \mathcal{S}, \mathcal{O}, \mathcal{A}, \pi\rangle$, determining the local properties of an agent in the system.
Herein, $\mathcal{S}$ denotes the set of the agent's local states, $\mathcal{O}$ is the set of possible local observations, ${\mathcal{A}}$ is the set of actions available to the agent, and $\pi:\mathcal{O}\times\mathcal{A}\rightarrow[0,1]$ is the agent's stochastic control policy. 
Based on this definition, the swarm MDP is constructed as $\langle N, \mathbb{A}, P, O, R\rangle$, where $N$ is the number of agents in the system and $\mathbb{A}$ is the aforementioned agent prototype.
The coupling of the agents is specified through a global state transition model $P: \mathcal{S}^N \times \mathcal{S}^N \times \mathcal{A}^N \rightarrow [0, \infty)$ and an observation model $O:\mathcal{S}^N\times\{1,\ldots,N\}\rightarrow\mathcal{O}$, which determines the local observation $\bm{o}^i\in\mathcal{O}$ for agent $i$ at a given swarm state $\bm{s}\in\mathcal{S}^N$, i.e., $\bm{o}^i=O(\bm{s},i)$.  
Finally, $R:\mathcal{S}^N\times\mathcal{A}^N\rightarrow\mathbb{R}$ is the global reward function, which encodes the cooperative task for the swarm by providing an instantaneous reward feedback $R(\bm{s},\bm{a})$ according to the current swarm state $\bm{s}$ and the corresponding joint action assignment $\bm{a}\in\mathcal{A}^N$ of the agents. 
The specific state dynamics and observation models considered in this paper are described in Section~\ref{sec:results}.

The model encodes two important properties of swarm networks: First, all agents in the system are assumed to be identical, and accordingly, they are all assigned the same decentralized policy~$\pi$. 
This is an immediate consequence of the two-step construction of the model, which implies that all agents share the same internal architecture. 
Second, the agents are only partially informed about the global system state, as prescribed by the observation model $O$. 
Note that both the transition model and the observation model are assumed to be invariant to permutations of the agents in order to ensure the homogeneity of the system.
For details, see \citep{sosic2017inverse}. 

\subsection{Local Observation Models}
The local observation $\bm{o}^i$ introduced in the last section is a combination of observations $o^i_\text{loc}$ an agent makes about local properties (like the agent's current velocity or its distance to a wall) and observations $O^i$ of other agents.
In order to describe the observation model used for the agents, we use an interaction graph representation of the swarm.
This graph is given by nodes $V=\{v_1, v_2, \dots, v_N \}$ corresponding to the agents in the swarm and an edge set $E \subset V \times V$, which we assume contains unordered pairs of the form $\{v_i, v_j\}$ indicating that agents $i$ and $j$ are neighbors.
The interaction graph is denoted as $\mathcal{G} = (V, E)$.
If both the set of nodes and the set of edges are not changing, we call $\mathcal{G}$ a static interaction graph; if either of the set undergoes changes, we instead refer to $\mathcal{G}$ as a dynamic interaction graph.

The set of neighbors of agent $i$ in the graph $\mathcal{G}$ is given by 
\begin{align*}
\mathcal{N}_\mathcal{G}(i) = \{j \given \{v_i, v_j\} \in E\}.
\end{align*}
Within this neighborhood, agent $i$ can sense local information about other agents, for example distance or bearing to each neighbor.
We denote the information agent $i$ receives from agent $j$ as $o^{i, j} = f(s^i, s^j)$, which is a function of the local states of agent $i$ and agent~$j$.
The observation $o^{i, j}$ is available for agent $i$ only if $j \in \mathcal{N}_\mathcal{G}(i)$. 
Hence, the complete state information agent $i$ receives from all neighbors is given by the set $O^i = \left\{o^{i, j} \given j \in  \mathcal{N}_\mathcal{G}(i) \right\}$.

As the observations of other agents are summarized in form of sets $\{O^i\}$, we require an efficient encoding that can be used as input to a neural network policy. 
In particular, it must meet the following two properties:
\begin{itemize}
 \item The encoding needs to be invariant to the indexing of the agents, respecting the unorderedness of the elements in the observation set. 
 Only by exploiting the system's inherent homogeneity we can escape the curse of dimensionality.
 \item The encoding must be applicable to varying set sizes because the local graph structure might change dynamically.
 Even if each agent can observe the entire system at all times, the encoding should be applicable for different swarm sizes. 
\end{itemize}

\subsection{Local Communication Models}

In addition to perceiving local state information of neighboring agents, the agents can also communicate information about the interaction graph $\mathcal{G}$ \citep{huettenrauch2018local}.
For example, agent $j$ can transmit the number of perceived neighbors to agent $i$.
Furthermore, the agents can also perform more complex operations on their local neighborhood graph.
For example, they could compute the shortest distance to a target point (such as an evader) that is perceived by at least one agent within their local sub-graph.
Hence, by using local communication protocols, observation $o^{i, j}$ can contain information about both, the local states $s^i$ and $s^j$ as well as the graph $\mathcal{G}$, i.e., $o^{i, j} = f(s^i, s^j, \mathcal{G})$.

\subsection{Mean Embeddings as State Representations for Swarms}
In the simplest case, the local observation $o^{i,j}$ that agent $i$ receives of agent $j$ is composed of the distance and the bearing angle of agent $i$ to agent $j$.
However, $o^{i,j}$ can also contain more complex information, such as relative velocities or orientations.
A straightforward way to represent the information set $O^i$ is to concatenate the local quantities $\{o^{i,j}\}_j$ into a single observation vector.
However, as mentioned before, this representation has various drawbacks as it ignores the permutation invariance inherent to a homogeneous agent network.
Furthermore, it grows linearly with the number of agents in the swarm and is, therefore, limited to a fixed number of neighbors when used in combination with neural network policies.

To resolve these issues, we treat the elements in the information set $O^i$ as samples from a distribution that characterizes the current swarm configuration, i.e., $o^{i,j} \sim p_i(\cdot \given \mathbf{s})$.
We can now use an empirical encoding of this distribution in order to achieve permutation invariance of the elements of $O^i$ as well as flexibility to the size of $O^i$.
As highlighted in Section~\ref{sec:mean_embs}, a simple way is to use a mean feature embedding, i.e., 
\begin{align*}
\hat{\mu}_{O^i} = \frac{1}{|O^i|} \sum_{o^{i,j} \in O^i} \phi(o^{i, j}),
\end{align*}
where $\phi$ defines the feature space of the mean embedding.
The input dimensionality to the policy is given by the dimensionality of the feature space of the mean embedding and, hence, it does not depend on the size of the information set $O^i$ any more.
This allows us to use the embedding $\hat{\mu}_{O^i}$ as input to a neural network used in deep RL.
In the following sections, we describe different feature spaces that can be used for the mean embedding.
Figure~\ref{fig:policy_models} illustrates the resulting policy architectures with further details given in Appendix~\ref{app:policy_archs}.

\begin{figure}
    \centering
    \begin{subfigure}[c]{0.55\columnwidth}
        \centering
        \includegraphics[width=\columnwidth]{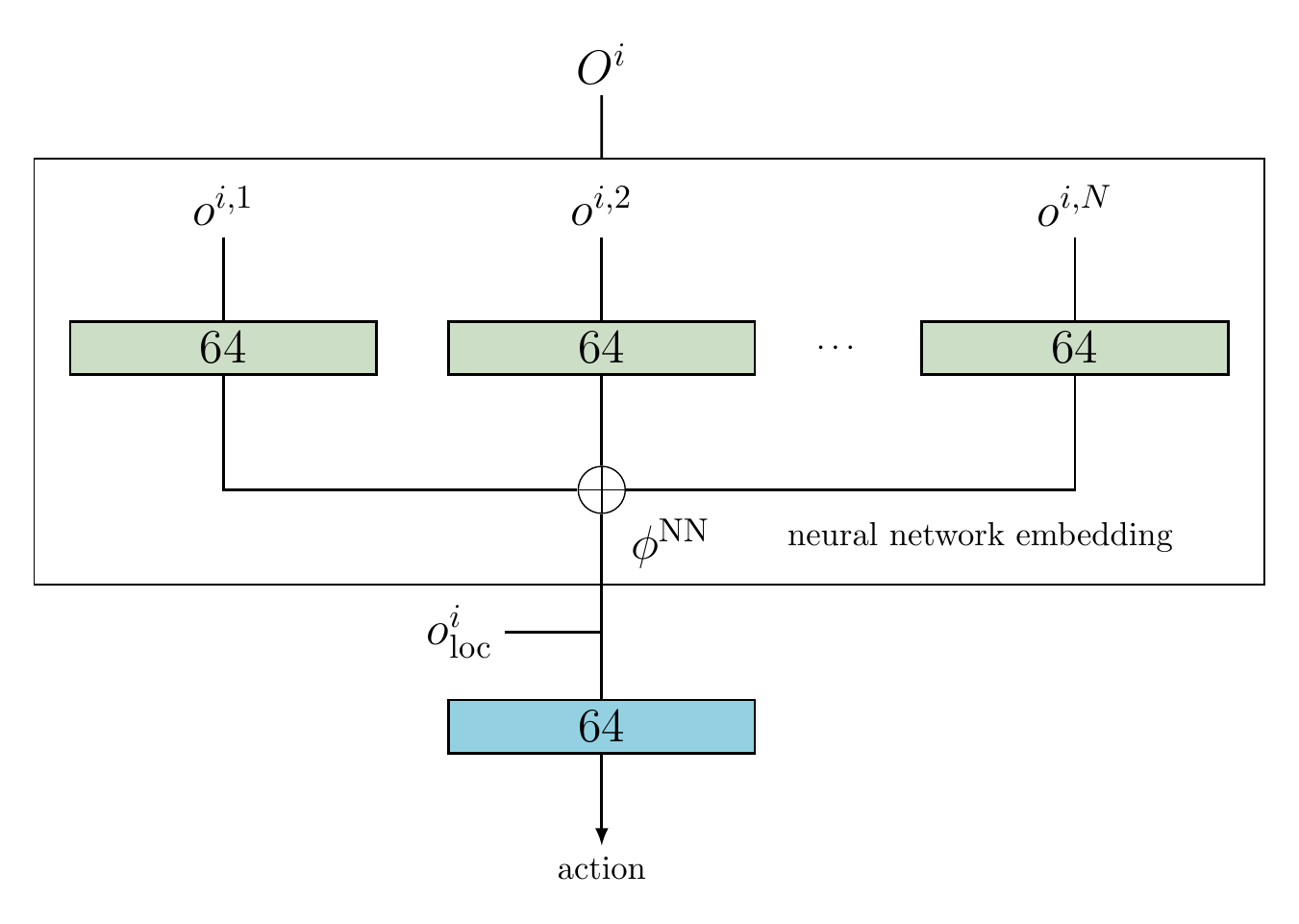}
        \caption{neural network embedding policy network}
        \label{fig:policy_model1}
    \end{subfigure}
    \hfill
    \begin{subfigure}[c]{0.17\columnwidth}
        \centering
        \includegraphics[width=\columnwidth]{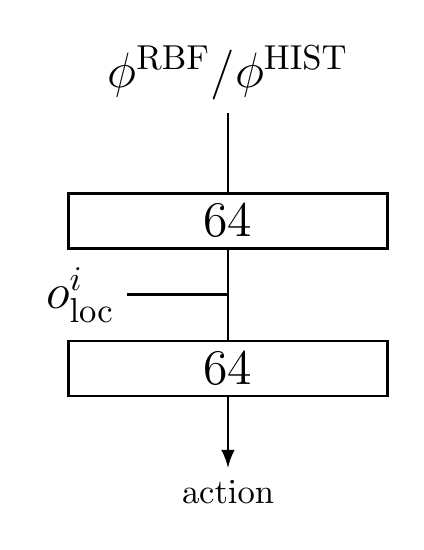}
        \caption{RBF and histogram embedding policy}
        \label{fig:policy_model2}
    \end{subfigure}
    \hfill
    \begin{subfigure}[c]{0.17\columnwidth}
        \centering
        \includegraphics[width=\columnwidth]{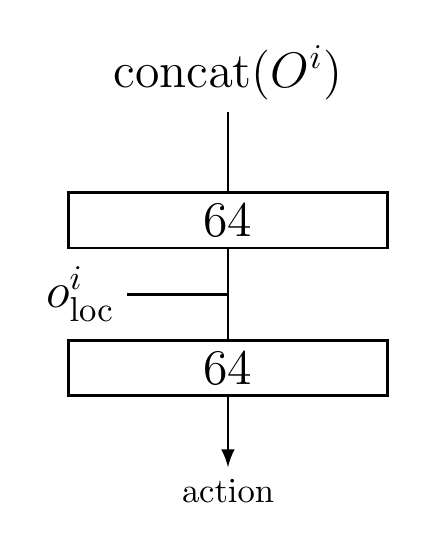}
        \caption{policy network for concatenation}
        \label{fig:policy_model3}
    \end{subfigure}
    \caption{Illustration of (a) the neural network mean embedding policy, (b) the network architecture used for the RBF and histogram representation, and (c) for the simple concatenation of observations. The numbers inside the boxes denote the dimensionalities of the hidden layers. The color coding in (a) highlights which layers share the same weights. The plus sign denotes the mean of the feature activations.}
    \label{fig:policy_models}
\end{figure}

\subsubsection{Neural Network Feature Embeddings}
In line with the deep RL paradigm, we propose to use a neural network as feature mapping~$\phi^{\textrm{NN}}$ whose parameters are determined by the reinforcement learning algorithm.
Using a neural network to define the feature space allows us to handle high dimensional observations, which is not feasible with traditional approaches such as histograms \citep{huettenrauch2018local}. 
In our experiments, a rather shallow architecture with one layer of RELU units already performed very well, but deeper architectures could be used for more complex applications.
To the best of our knowledge, we present the first approach for using neural networks to define the feature space of a mean embedding.

\subsubsection{Histograms}
An alternative feature space are provided by histograms, which can be related to image-like representations.
In this approach, we discretize the space of certain features, such as the distance and bearing to other agents, into a fixed number of bins.
This way, we can collect information about neighboring agents in the form of a fixed-size multi-dimensional histogram.
Herein, the histogram bins define a feature mapping $\phi^{\textrm{HIST}}$ using a one-hot-coding for each observed agent.
A detailed description of this approach can be found in our previous work \citep{huettenrauch2018local}.
While the approach works well in discrete environments where each cell is only occupied by a single agent, the representation can lead to blurring effects between agents in the continuous case. 
Moreover, the histogram approach does not scale well with the dimensionality of the feature space.

\subsubsection{Radial Basis Functions}
A specific problem of the histogram approach is the hard assignment of agents into bins, 
which results in abrupt changes in the observation space when a neighboring agent moves from one bin to another.
A more fine-grained representation can be achieved by using RBF networks with a fixed number of basis functions evenly distributed over the observation space. 
The resulting feature mapping $\phi^{\textrm{RBF}}$ is then defined by the activations of each basis function and can be seen as a ``soft-assigned'' histogram.
However, both representations (histogram and RBF) suffer from the curse of dimensionality, as the number of required basis functions typically increases exponentially with the number of dimensions of the observation vector.

\subsection{Other Representation Techniques}
\label{sec:repr_techs}
Inspired by the work of \citet{mordatch2018emergence}, we also investigate a policy function that uses a softmax pooling layer instead of the mean embedding.
The elements of the pooling layer $\psi = \left[\psi_1, \dots, \psi_K\right]$ are given by
\begin{align*}
\psi_k = \frac{\sum_{o^{i,j} \in O^i} \exp \left(\alpha \phi_k(o^{i, j}) \right) \phi_k(o^{i, j})}{\sum_{o^{i,j} \in O^i} \exp \left(\alpha \phi_k(o^{i, j}) \right)}
\end{align*}
for each feature dimension of $\phi = \left[ \phi_1, \dots, \phi_K \right]$ with a temperature parameter $\alpha$.
Note that the representation becomes identical to our mean embedding for $\alpha = 0$, while setting $\alpha \gg 1$ results in max-pooling and $\alpha \ll -1$ corresponds to min-pooling.
In our experiments, we choose $\alpha=1$ as a trade-off between a mean embedding and max-pooling and additionally study the performance of max-pooling over each individual feature dimension.

\subsection{Adaption of TRPO to the Homogeneous Swarm Setup}
\citet{gupta2017cooperative} present a parameter-sharing variant of TRPO that can be used in a multi-agent setup.
During the learning phase, the algorithm collects experiences made by all agents and uses these experiences to optimize one policy with a single set of parameters $\theta$.
Since, in the swarm setup, we assume homogeneous agents that are potentially indistinguishable to each other, we omit the agent index introduced by \cite{gupta2017cooperative}.
The optimization problem is expressed using advantage values based on all agents' observations.
During execution, however, each agent has only access to its own perception.
Hence, the terminology of centralized--learning\textslash decentralized--execution is chosen.

During the trajectory roll-outs, we use a sub-sampling strategy to achieve a trade-off between the number of samples and the variability in advantage values seen by the learning algorithm.
Our implementation is based on the OpenAI baselines version of TRPO with 10 MPI workers, where each worker samples 2048 time steps, resulting in $2048 N$ samples.
Subsequently, we randomly choose the data of 8 agents, yielding $2048 \times 10 \times 8 = 163840$ samples per TRPO iteration.
The chosen number of samples worked well throughout our experiments and was not extensively tuned.

\section{Experimental Results}
\label{sec:results}
Our experiments are designed to study the use of mean embeddings in a cooperative swarm setting.
The three main aspects are:
\begin{enumerate}
	\item How do the different mean embeddings (neural networks, histograms and RBF representation) compare when provided with the same state information content?
	\item How does the mean embedding using neural networks perform when provided with additional state information while keeping the dimensionality of the feature space constant?
	\item How does the mean embedding of neural network features compare against other pooling techniques?
\end{enumerate}
In this section, we first introduce the swarm model used for our experiments and present the results of different evaluations afterwards.
During a policy update, a fixed number of $K$ trajectories are sampled, each yielding a return of $G_k = \sum_{t=1}^{T} r(t)$.
The results are presented in terms of the average return,  denoted as $\bar{G} = \frac{1}{K} \sum_{k=1}^{K} G_k$.
Videos demonstrating the agents' behavior in the different tasks can be found online.\footnote{\url{http://computational-learning.net/deep_rl_for_swarms}} 

\subsection{Swarm Models}
Our agents are modeled as unicycles \citep[a commonly used agent model in mobile robotics; see, for example,][]{egerstedt2001formation}, where the control parameters either manipulate the linear and angular velocities $v$ and $\omega$ (single integrator dynamics) or the corresponding accelerations $\dot{v}$ and $\dot{\omega}$ (double integrator dynamics).
In the single integrator case, the state of an agent is defined by its location $\bm{x} = (x, y)$ and orientation $\phi$.
In case of double integrator dynamics, the agent is additionally characterized by its current velocities. 
The exact state definition and kinematic models can be found in Appendix~\ref{sec:app_kinematics}.
Note that these agent models are more complex than what is typically considered in optimization-based approaches, which mostly assume single integrator dynamics directly on $\bm{x}$. 
Depending on the task, we either opt for a closed state space where the limits act as walls, or a periodic toroidal state space where agents exceeding the boundaries reappear on the opposite side of the space.
Either way, the state is bounded by $x_\text{max} = y_\text{max} = 100$.

We study two different observation scenarios for the agents, i.e., global observability and local observability. In the case of global observability, 
all agents are neighbors, i.e.
\begin{align*}
\mathcal{N}_\mathcal{G}(i) = \{j \in \{1, \dots, N\} \given i \neq j\},
\end{align*}
which corresponds to a fully connected static interaction graph.
For the local observability case, we use $\Delta$-disk proximity graphs, where edges are formed if the distance $d^{i, j} =\sqrt{(x^i -x^j)^2 + (y^i - y^j)^2}$ between agents $i$ and $j$ is less than a pre-defined cut-off distance~$d_c$ for communication, resulting in a dynamic interaction graph. 
The neighborhood set of the graph is then defined as
\begin{align*}
\mathcal{N}_{\mathcal{G}}(i) = \{j \in \{1, \dots, N\} \given i \neq j, ~ d^{i, j} \leq d_c\}.
\end{align*}
For a detailed description of all observational features available to the agents in the tasks, see Appendices \ref{sec:app_obs_model} and \ref{sec:app_task_obs_model}.

\subsection{Rendezvous}
In the rendezvous problem, the goal is to minimize the distances between all agents.
The reason why we choose this experiment is because a simple optimization-based baseline controller can be defined by the consensus protocol,
\begin{align*}
\dot{\bm{x}}^i = - \sum_{j \in \mathcal{N}(i)} (\bm{x}^i - \bm{x}^j),
\end{align*}
where $\bm{x}^i = (x^i, y^i)$ denotes the location of agent $i$.
To make the solution compatible to the double integrator agent model, we make use of a PD-controller (see Appendix \ref{sec:app_kinematics} for details).
The reward function for the problem can be found in Appendix \ref{appendix:rew_rend}.

\begin{figure}
	\centering
	\includegraphics[width=0.75\columnwidth]{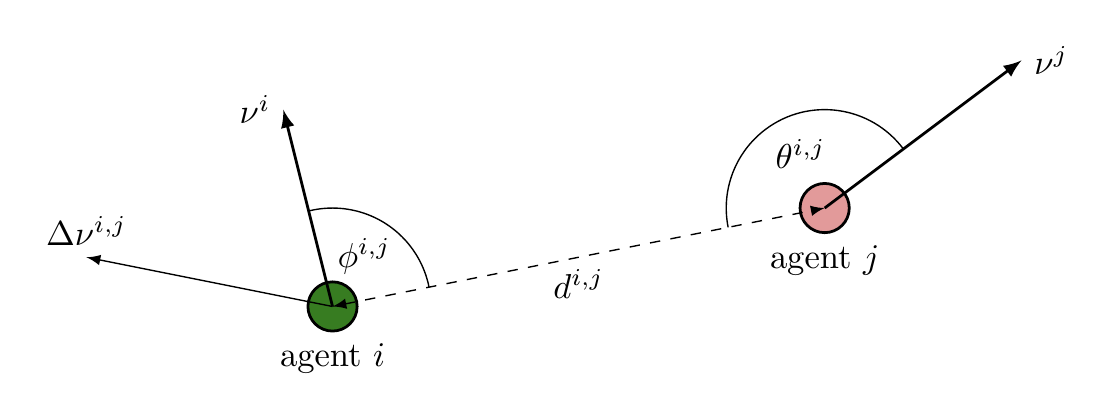}
	\caption{%
		Illustration of two neighboring agents facing the direction of their velocity vectors $\nu^i$ and $\nu^j$, along with the observed quantities, shown with respect to agent~$i$.
	The observed quantities are the bearing $\phi^{i,j}$ to agent $j$, agent $j$'s relative orientation~$\theta^{i,j}$ to agent $i$, their distance $d^{i,j}$ and a relative velocity vector $\Delta \nu^{i,j} = \nu^{i} - \nu^{j}$.
	In this trivial example, agent $i$'s observed neighborhood size as well as the neighborhood size communicated by agent $j$ are $\lvert \mathcal{N}(i) \rvert = \lvert \mathcal{N}(j) \rvert = 1$.
	}
	\label{fig:prop_vis}
\end{figure}

We evaluate different observation vectors $o^{i,j}$ which are fed into the policy.
To compare the histogram and RBF embedding with the proposed neural network approach, we restrict the \emph{basic} observation model (see below) to a set of two features: 
the distance $d^{i, j}$  between two agents and the corresponding bearing~$\phi^{i, j}$.
This restriction allows for a comparison to the optimization-based consensus protocol, which is based on displacements (an equivalent formulation of distance and bearing).
To show that the neural network embeddings can be used with more informative observations, we further introduce an \emph{extended} set and a communication (\emph{comm}) set.
These sets may include relative orientations $\theta^{i, j}$ or relative velocities $\Delta \nu^{i, j}$ (depending on the agent dynamics), as well as the own neighborhood size and those of the neighbors.
An illustration of these quantities can be found in Figure \ref{fig:prop_vis}.

\subsubsection{Global Observability}
\begin{figure}[t]
	\centering
	\captionsetup[subfigure]{justification=centering, margin={1.1cm,0cm}}
	\subcaptionbox{20 agents with global observability \label{fig:rend_20a_glob_obs_top5_median}}{%
		\includegraphics[scale=1]{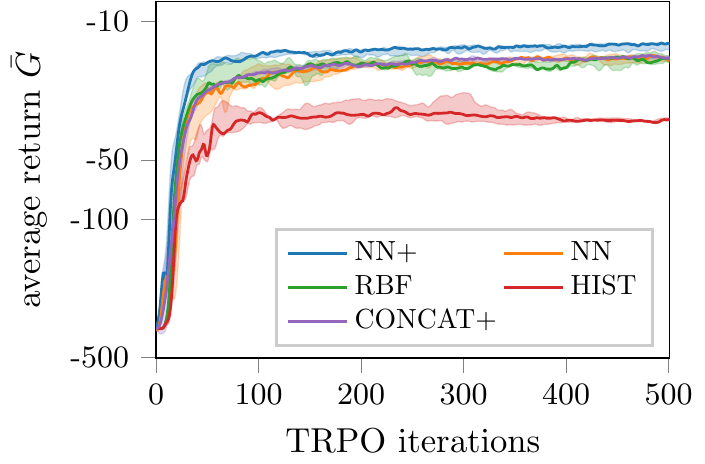}%
	}
	\subcaptionbox{20 agents with local observability \label{fig:rend_20a_40comm_top5_median}}{%
		\includegraphics[scale=1]{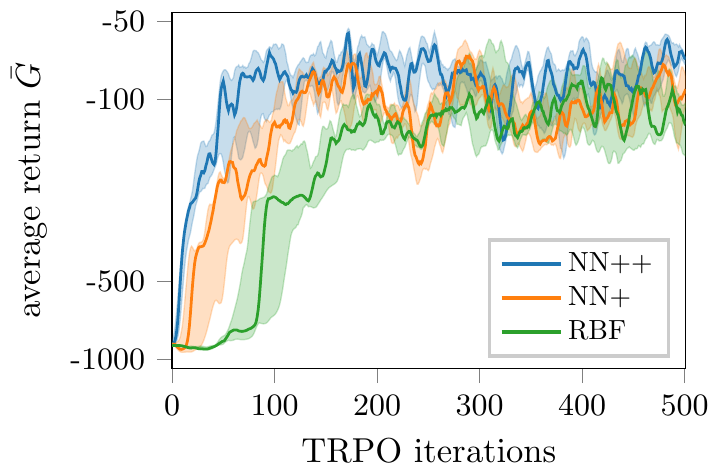}%
	}
	\caption{Learning curves for the rendezvous task with different observation models. The curves show the median of the average return $\bar{G}$ based on the top five trials on a log scale. \textbf{Legend:} NN++: neural network mean embedding of \emph{comm} set, NN+: neural network mean embedding of \emph{extended} set, NN: neural network embedding of \emph{basic} set, RBF: radial basis function embedding of \emph{basic} set, HIST: histogram embedding of \emph{basic} set, CONCAT+: simple concatenation of \emph{extended} set.}
	\label{fig:rend_lcs}
\end{figure}

First, we study the rendezvous problem with 20 agents in the global observability setting with double integrator dynamics to illustrate the algorithm's ability to handle complex dynamics.
To this end, we compare the performances of policies using histogram, RBF and neural network embeddings on the \emph{basic} set, as well as neural network embeddings on the \emph{extended} set. The observations $o^{i,j}$ in the \emph{basic} set comprise the distance $d^{i, j}$ and bearing $\phi^{i, j}$.
In the \emph{extended} set, which is processed only via neural network embeddings, we additionally add neighboring agents' relative orientations $\theta^{i, j}$ and velocities $\Delta \nu^{i,j}$.
The local properties $o^i_\text{loc}$ consist of a shortest distance and orientation to the closest boundary, i.e., $d^i_\text{wall}$ and $\phi^i_\text{wall}$.
The sets are summarized as follows:
\begin{align*}
	\emph{Basic}: \quad o^{i, j} &= \{d^{i, j},~ \phi^{i, j}\} & o^i_\text{loc} &= \{d^i_\text{wall},~ \phi^i_\text{wall}\}\\
	\emph{Extended}: \quad o^{i, j} &= \{d^{i, j},~ \phi^{i, j},~ \theta^{i, j},~ \Delta \nu^{i,j} \} & o^i_\text{loc} &= \{d^i_\text{wall},~ \phi^i_\text{wall}\}.
\end{align*}
The results are shown in Figure \ref{fig:rend_20a_glob_obs_top5_median}.
On first sight, they reveal that all shown methods eventually find a successful strategy, with the histogram approach showing worst performance.
Upon a closer look, it can be seen that the best solutions are found with the neural network embedding, in which case the learning algorithm also converges faster, demonstrating that this form of embedding serves as a suitable representation for deep RL.
However, there are two important things to note:
\begin{itemize}
	\item 
	The differences between the approaches seem to be small due to the wide range of obtained reward values, but the NN+ method brings in fact a significant performance gain. 
	Compared to the NN and RBF embedding, the performance of the learned NN+ policy is ${\sim}10 \%$ better in terms of the average return of an episode (Figure~\ref{fig:rend_20a_glob_obs_top5_median}) and almost twice as good (${\sim}\num{4e-2}$ versus ${\sim}\num{8e-2}$) in terms of the mean distance between agents at the steady state solution after around 200 time steps (Figure~\ref{fig:rend_20a_20a_glob_obs_dist}). Furthermore, the NN+ embedding reaches the mean distance achieved by the NN and RBF embeddings roughly 20 to 30 time steps earlier, which corresponds to an improvement of ${\sim}25 \%$.	
	\item Although the performance gain of NN+ can be partly explained by the use of the extended feature set, experiments with the same feature set using the histogram\textslash RBF approach did \textit{not} succeed to find solutions to the rendezvous problem; hence, the corresponding results are omitted. The reason is that the dimensionality of the input space scales exponentially for the histogram\textslash RBF approach while only linearly for the neural network embedding, which results in a more compact feature representation that keeps the learning problem tractable. 
\end{itemize}
 Together, these two observations suggest that the neural network embedding provides a suitable learning architecture for deep RL, whereas the histogram\textslash RBF approach is only suited for low-dimensional spaces.

\begin{figure}[t]
	\begin{subfigure}[c]{0.45\columnwidth}
		\centering
		\begin{subfigure}[c]{0.49\columnwidth}
			\centering
			\includegraphics[scale=1]{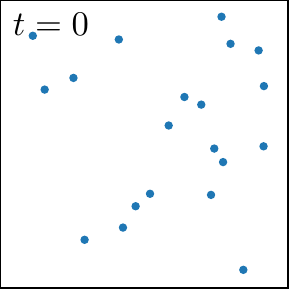}
		\end{subfigure}
		\begin{subfigure}[c]{0.49\columnwidth}
			\centering
			\includegraphics[scale=1]{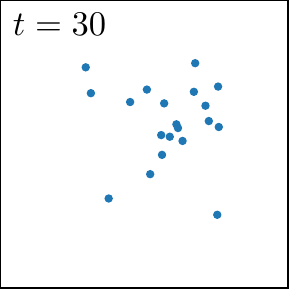}
		\end{subfigure}
		
		\vspace{1.25em}
		
		\begin{subfigure}[c]{0.49\columnwidth}
			\centering
			\includegraphics[scale=1]{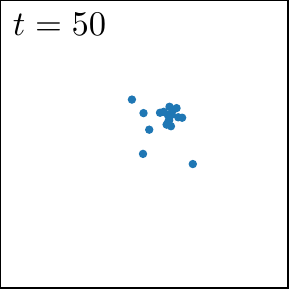}
		\end{subfigure}
		\begin{subfigure}[c]{0.49\columnwidth}
			\centering
			\includegraphics[scale=1]{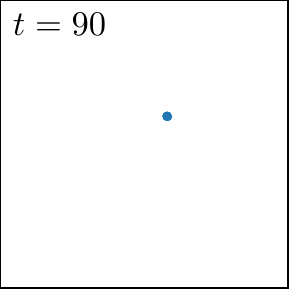}
		\end{subfigure}
		\caption{snapshots}
	\end{subfigure}
	\hfill
	\begin{subfigure}[c]{0.5\columnwidth}
		\centering
		\includegraphics[scale=0.8]{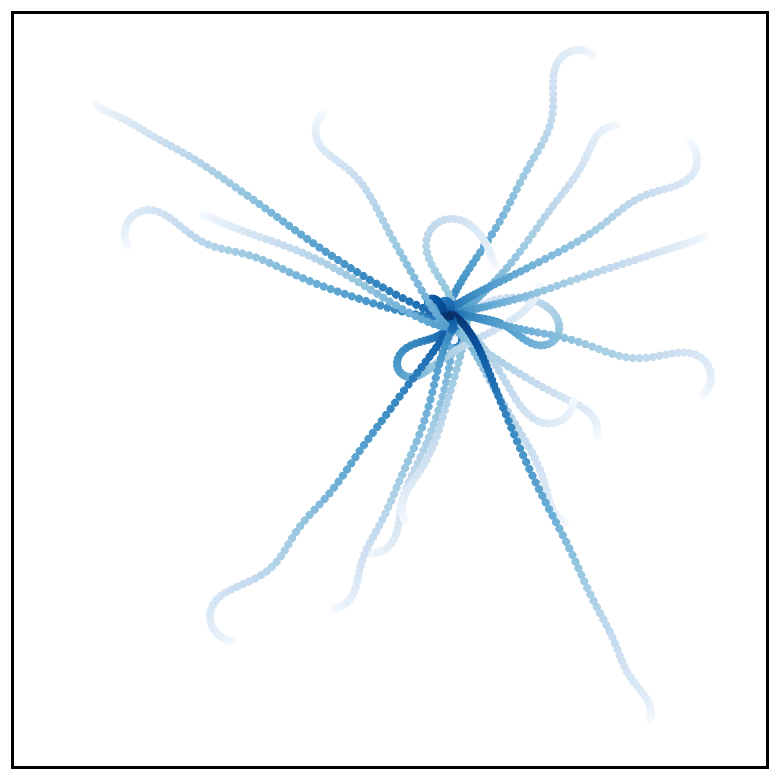}
		\caption{full episode}
		\label{fig:rend_20a_vis3}
	\end{subfigure}
	
	\caption{Visualization of a learned policy for the pursuit evasion task. The policy is learned and executed by 10 agents using a neural network mean embedding of the \emph{extended} set. Pursuers are illustrated in blue, the evader is highlighted in red. Visualization of a learned policy for the rendezvous task. The policy is learned and executed by 20 agents using a neural network mean embedding of the \emph{extended} set. }
	\label{fig:rend_20a_vis}
\end{figure}

Figure \ref{fig:rend_20a_vis} shows a visualization of a policy using the neural network mean embedding of the \emph{extended} set. After random initialization, the agents' locations quickly converge to a single point.
Figure \ref{fig:rend_mean_dist_comp} shows performance evaluations of the best policies found with each of the mean embedding approaches.
We plot the evolution of the mean distance between all agents over 1000 episodes with equal starting conditions.
We also include the performance of the PD-controller defined in Appendix \ref{sec:app_kinematics}.
It can be seen in Figures~\ref{fig:rend_20a_20a_glob_obs_dist} and \ref{fig:rend_20a_20a_dist} that the policies using the neural network embeddings decrease the mean distance most quickly and also find the best steady-state solutions among all learning approaches. 
While the optimization-based solution (PD) eventually drives the mean distance to zero, a small error remains for the learning-based approaches.
However, the learned policies are faster in reducing the distance and therefore show a better average reward.
Although the optimization-based policy is guaranteed to find an optimal stationary solution, the approach is build for simpler dynamics and hence performs suboptimally in the considered scenario.
Note, that the controller gains for this approach have been tuned manually to maximize performance. 

In order to show the generalization abilities of the embeddings, we finally evaluate the obtained policies (except for the concatenation) with 100 agents.
The results are displayed in Figure \ref{fig:rend_20a_100a_dist}.
Again, the neural network embedding of the \emph{extended} set is quickest in reducing the inter-agent distances, resulting in the best overall performance.

\subsubsection{Local Observability}
The local observability case is studied with 20 agents and a communication cut-off distance of $d_c=40$.
Due to the increased difficulty of the task, we resort to single integrator dynamics for this experiment.
Again, we evaluate the \emph{basic} and the \emph{extended} set, which in this case contains the single integrator state information.
Accordingly, we remove the relative velocities from the information sets.
Moreover, we employ a local communication strategy that transmits the number of observed neighbors as additional information.
Note that this information can be used by the agents to estimate in which direction the center of mass of the swarm is located.

While the received neighborhood sizes $\{\lvert \mathcal{N}(j) \rvert\}_{j \in \mathcal{N}(i)}$ are treated as part of agent $i$'s local observation of the swarm, the own perceived neighborhood size $\lvert \mathcal{N}(i) \rvert$ is considered as part of the local features $o^i_\text{loc}$.
The observation models for the local observability case are thus summarized as:
\begin{align*}
	\emph{Basic}: \quad o^{i, j} &= \{d^{i, j},~ \phi^{i, j}\} & o^i_\text{loc} &= \{d^i_\text{wall},~ \phi^i_\text{wall}\}\\
	\emph{Extended}: \quad o^{i, j} &= \{d^{i, j},~ \phi^{i, j},~ \theta^{i, j} \} & o^i_\text{loc} &= \{d^i_\text{wall},~ \phi^i_\text{wall}\} \\
	\emph{Comm}: \quad o^{i, j} &= \{d^{i, j},~ \phi^{i, j},~ \theta^{i, j},~ \lvert \mathcal{N}(j) \rvert \} &  o^i_\text{loc} &= \{d^i_\text{wall},~ \phi^i_\text{wall},~ \lvert \mathcal{N}(i) \rvert \}.
\end{align*}

For the experiment, we limit our comparison to RBF embeddings (which showed best performance among all non-neural-network solutions) of the \emph{basic} set and neural network embeddings of the \emph{extended} set and the \emph{comm} set.
The results are illustrated in Figure~\ref{fig:rend_20a_40comm_top5_median}, which shows that the neural network embeddings lead to a quicker learning progress. Furthermore, by introducing the \emph{comm} model, a higher return is achieved.
Compared to the global observability case, however, the learning process exhibits an increased variance caused by the information loss in the reward signal (see Appendix~\ref{appendix:rewardFunctions}).

Figure \ref{fig:rend_20a_20a_dist} illustrates the performances of the learned policies.
Again, the neural network embedding is quicker in reducing the inter-agent distances and converges to better steady-state solutions.
In order to test the efficacy of the communication protocol, we further evaluate the learned policies with 10 agents.
The results are displayed in Figure \ref{fig:rend_20a_10a_dist}.
As expected, the performance decreases due to the lower chance of agents seeing each other but we still notice a benefit caused by the communication.

\begin{figure}[t]
	\subcaptionbox{20 agents (global observability)\label{fig:rend_20a_20a_glob_obs_dist}}{%
		\includegraphics[scale=1]{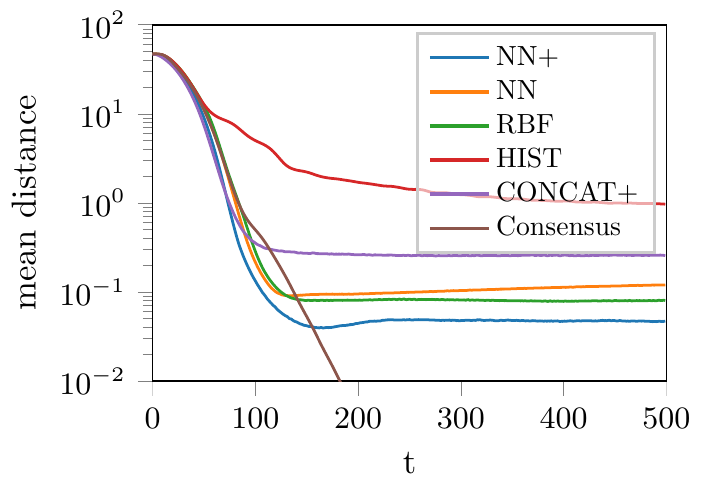}%
	}
	\hfill
	\subcaptionbox{100 agents (global observability) \label{fig:rend_20a_100a_dist}}{%
		\includegraphics[scale=1]{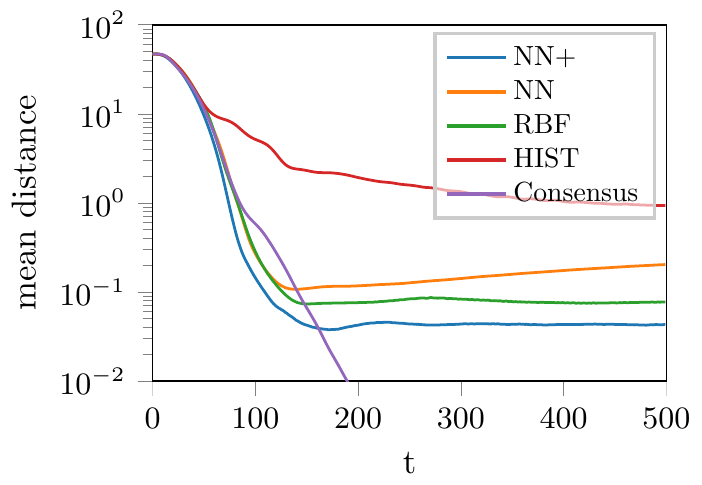}%
	}
	
	\vspace{1em}
	
	\subcaptionbox{20 agents (local observability)\label{fig:rend_20a_20a_dist}}{%
		\includegraphics[scale=1]{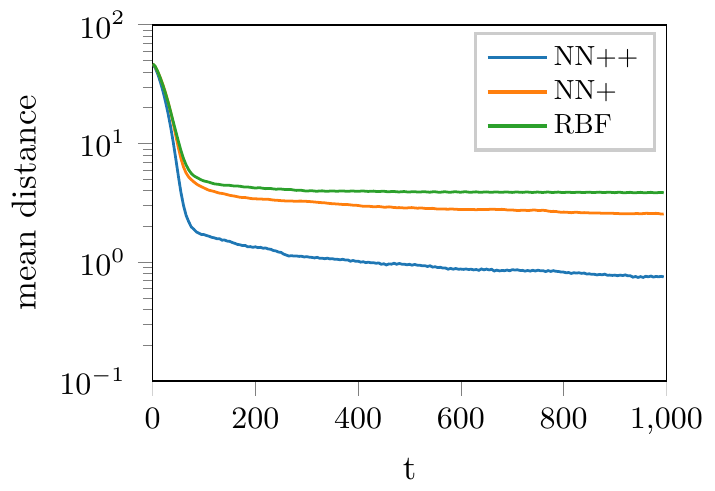}%
	}
	\hfill
	\subcaptionbox{10 agents (local observability)\label{fig:rend_20a_10a_dist}}{%
		\includegraphics[scale=1]{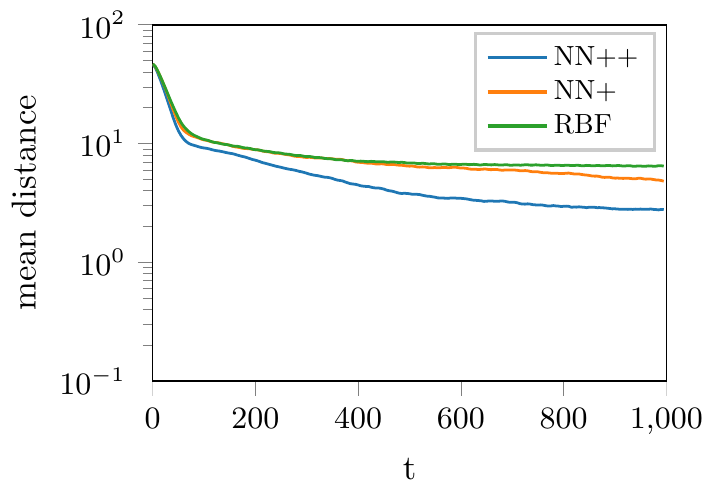}%
	}
	\caption{Comparison of the mean distance between agents in the rendezvous experiment achieved by the best learned policies and the consensus protocol. In (a) and (b), the policy is learned with 20 agents and executed by 20 and 100 agents, respectively. In (c) and (d), the policy is learned with 20 agents and executed by 20 and 10 agents. Results are averaged over 1000 episodes with identical starting conditions.}
	\label{fig:rend_mean_dist_comp}	
\end{figure}

\subsection{Pursuit Evasion with a Single Evader}
Our implementation of the pursuit evasion scenario is based on the work by \cite{zhou2016cooperative}, from which we adopt the evader strategy.
The strategy is based on Voronoi regions, which the pursuers try to minimize and the evader tries to maximize.
While the original paper considers a closed world, we change the world type from closed to periodic, thereby making it impossible to trap the evader in a corner.
In order to encourage a higher level of coordination between the agents, we set the evader's maximum velocity to twice the pursuers' maximum velocity.
An episode ends once the evader is caught, i.e., if the distance of the closest pursuer is below a certain threshold.
In all our experiments, the evader policy is fixed and not part of the learning process.
The reward function for the problem is based on the shortest distance of the closest pursuer and can be found in Appendix \ref{appendix:rew_pe}.

\subsubsection{Global Observability}
\label{sec:evasionGlobalObs}
Again, we study the global observability case with ten agents.
Since the pursuit of an evader is a more challenging task already, we reduce the movement complexity to single integrator dynamics.
The \emph{basic} and \emph{extended} set are equal to those in the rendezvous experiment with single integrator dynamics, with additional information about the evader in the local properties $o^i_\text{loc}$.
In here, we add the distance $d^{i, e}$ and bearing $\phi^{i, e}$ of agent $i$ to the evader $e$.
Accordingly, the observation sets are given as:
\begin{align*}
	\emph{Basic}: \quad o^{i, j} &= \{d^{i, j},~ \phi^{i, j}\} & o^i_\text{loc} &= \{d^i_\text{wall},~ \phi^i_\text{wall},~ d^{i, e},~ \phi^{i, e}\} \\
	\emph{Extended}: \quad o^{i, j} &= \{d^{i, j},~ \phi^{i, j},~ \theta^{i,j} \} & o^i_\text{loc} &= \{d^i_\text{wall},~ \phi^i_\text{wall},~ d^{i, e},~ \phi^{i, e}\}.
\end{align*}

\begin{figure}
	\centering
	\captionsetup[subfigure]{justification=centering, margin={1.1cm,0cm}}
	\subcaptionbox{10 agents with global observability \label{fig:pe_10a_full_obs_median}}{%
		\includegraphics[scale=1]{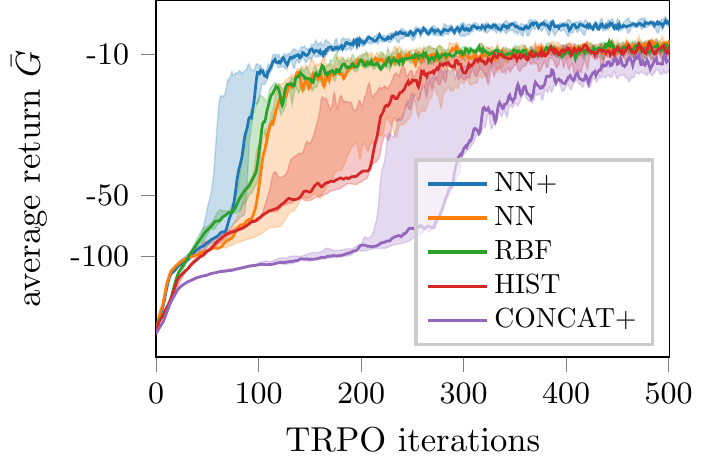}%
	}
	\subcaptionbox{20 agents with local observability \label{fig:pe_20a_40comm_top5_median}}{%
		\includegraphics[scale=1]{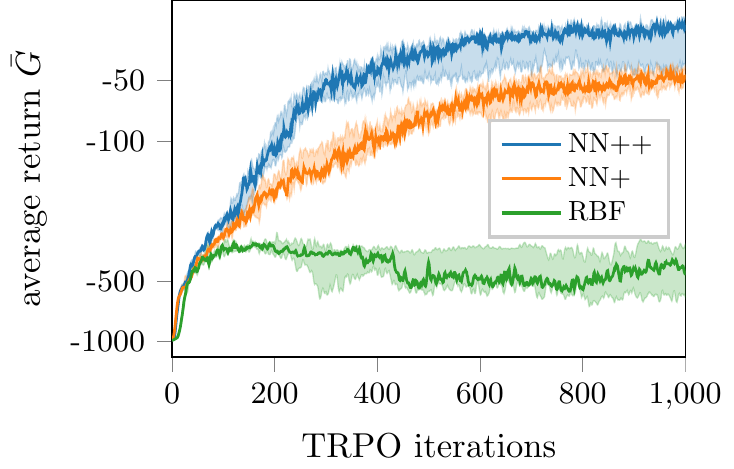}%
	}
	\caption{Learning curves for the pursuit evasion task with different observation models. The curves show the median of the average return $\bar{G}$ based on the top five trials on a log scale. \textbf{Legend:} NN++: neural network mean embedding of \emph{comm} set, NN+: neural network mean embedding of \emph{extended} set, RBF: radial basis function embedding of \emph{basic} set, HIST: histogram embedding of \emph{basic} set, CONCAT+: concatenation of \emph{extended} set.}
	\label{fig:pe_lcs}
\end{figure}

The results in Figure \ref{fig:pe_10a_full_obs_median} reveal that successful strategies can be obtained with all methods.
However, this time, a clear advantage can be seen for the policies using neural network mean embeddings of the \emph{extended} set, both in terms of behavior quality and in the number of samples necessary to find the solution.

\begin{figure}
	\begin{subfigure}[c]{0.45\columnwidth}
		\centering
		\begin{subfigure}[c]{0.49\columnwidth}
			\centering
			\includegraphics[scale=1]{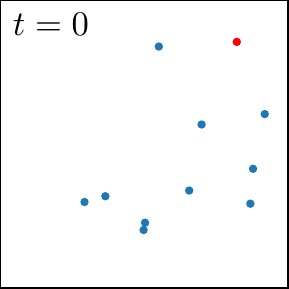}
		\end{subfigure}
		\begin{subfigure}[c]{0.49\columnwidth}
			\centering
			\includegraphics[scale=1]{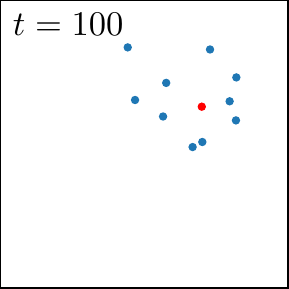}
		\end{subfigure}
		
		\vspace{1.25em}

		\begin{subfigure}[c]{0.49\columnwidth}
			\centering
			\includegraphics[scale=1]{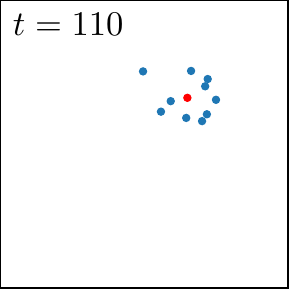}
		\end{subfigure}
		\begin{subfigure}[c]{0.49\columnwidth}
			\centering
			\includegraphics[scale=1]{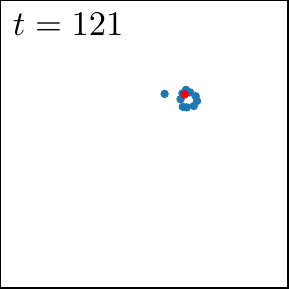}
		\end{subfigure}
		\caption{snapshots}
	\end{subfigure}
	\hfill
	\begin{subfigure}[c]{0.5\columnwidth}
		\centering
		\includegraphics[scale=0.8]{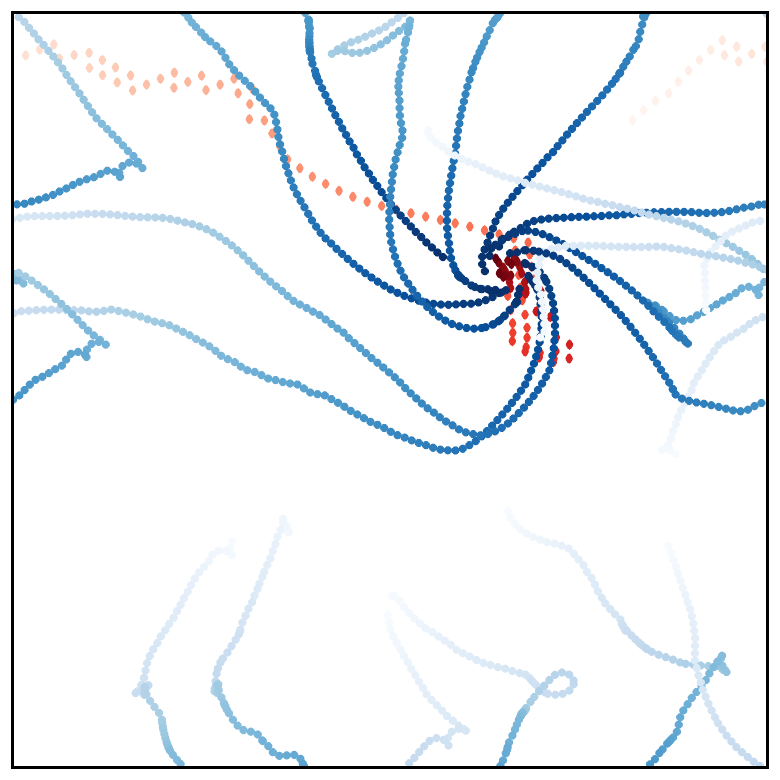}
		\caption{full episode}
		\label{fig:pe_10a_vis3}
	\end{subfigure}
	
	\caption{Visualization of a learned policy for the pursuit evasion task. The policy is learned and executed by 10 agents using a neural network mean embedding of the \emph{extended} set. Pursuers are illustrated in blue, the evader is highlighted in red.}
	\label{fig:pe_10a_vis}
\end{figure}

Figure \ref{fig:pe_10a_vis} illustrates the strategy that such a policy  exerts.
After random initialization, the agents first spread in a way that leaves no possibility for the evader to increase its Voronoi region, thereby keeping the evader almost on the same spot.
Once this configuration is reached, they surround the evader in a circular pattern and start to reduce the distance until one pursuer successfully reaches the distance  threshold.

\begin{figure}[t]
	\centering
	\begin{subfigure}[t]{0.49\columnwidth}
		\includegraphics[scale=1]{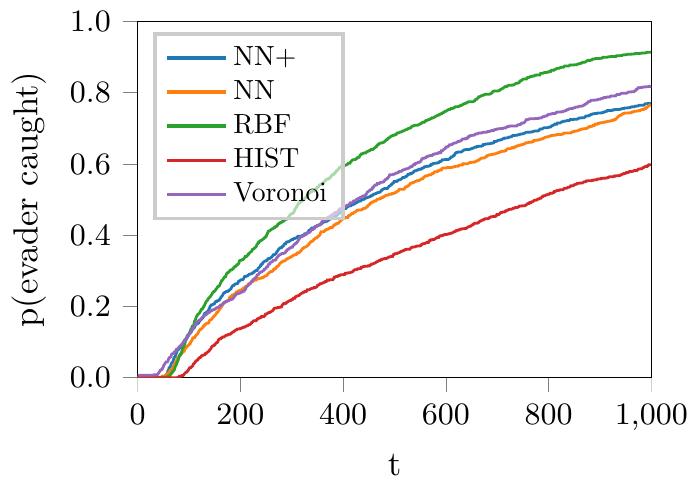}
		\caption{5 agents}
		\label{fig:pe_10a_5a_perf}
	\end{subfigure}
	\begin{subfigure}[t]{0.49\columnwidth}
		\includegraphics[scale=1]{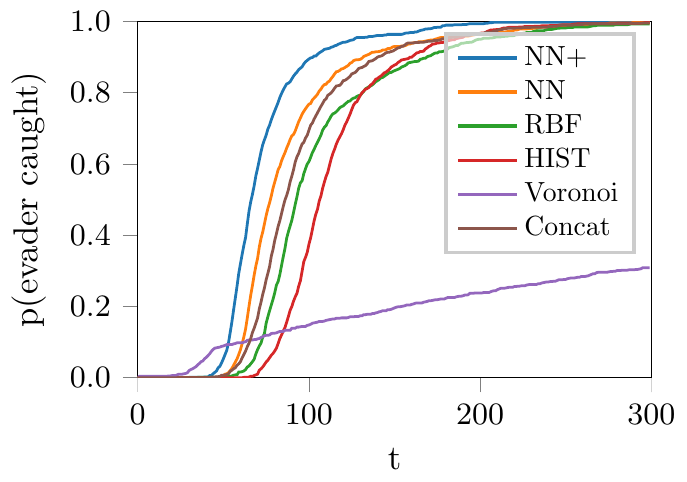}
		\caption{10 agents}
		\label{fig:pe_10a_10a_perf}
	\end{subfigure}
	
	\begin{subfigure}[t]{0.49\columnwidth}
		\includegraphics[scale=1]{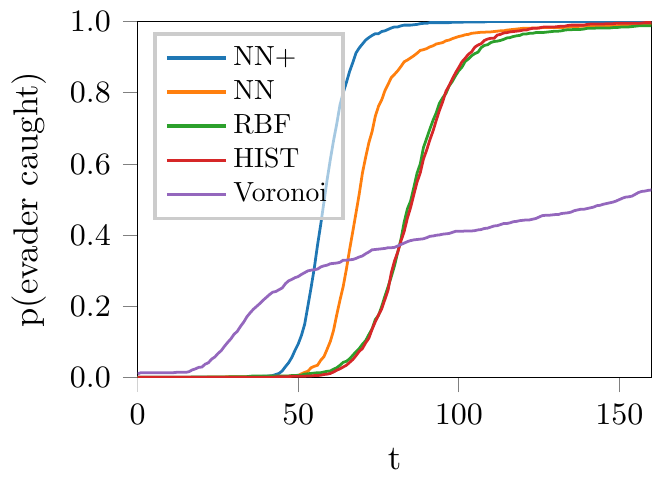}
		\caption{20 agents}
		\label{fig:pe_10a_20a_perf}
	\end{subfigure}
	\begin{subfigure}[t]{0.49\columnwidth}
		\includegraphics[scale=1]{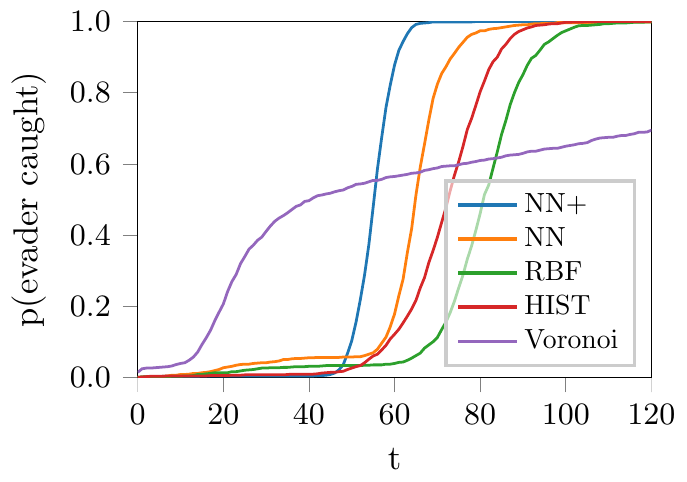}
		\caption{50 agents}
		\label{fig:pe_10a_50a_perf}
	\end{subfigure}
	\caption{Performance comparison of the best learned policies and the optimization approach minimizing Voronoi regions in the pursuit evasion task with global observability. The curves show the probability that the evader is caught after $t$ time steps. All policies are learned with 10 agents but executed with different agent numbers, as indicated below each subfigure. Results are averaged over 1000 episodes with identical starting conditions.
	}
	\label{fig:pe_perf_comp}
\end{figure}

To investigate the performance of the best mean embedding policies (learned with 10 agents), we estimate the corresponding probabilities that the evader is caught within a certain time frame.
For the sake of completeness, we also include the method proposed by \cite{zhou2016cooperative}, which was originally not designed for a setup with a faster evader, though.
The results are plotted in Figure \ref{fig:pe_perf_comp} as the fraction of episodes ending at the respective time instant, averaged over 1000 episodes.
The plot in Figure \ref{fig:pe_10a_10a_perf} reveals that the evader may be caught using all presented methods if the policies are executed for long time periods.
As already indicated by the learning curves, using a neural network mean embedding representation yields the quickest capture among all methods.
The additional information in the \emph{extended} set further increases performance.

Next, we examine the generalization abilities of the learned policies, this time on scenarios with 5, 20 and 50 agents (Figures~\ref{fig:pe_10a_5a_perf},\subref{fig:pe_10a_20a_perf},\subref{fig:pe_10a_50a_perf}).
Increasing the amount of agents leads to a quicker capture for all methods; however, the best performance is still shown by the agents executing a neural network policy based on embeddings of the \emph{extended} set.
Interestingly, when using fewer agents than in the original setup (Figure~\ref{fig:pe_10a_5a_perf}), all methods struggle to capture the evader.
After inspection of the behavior, we found that the strategy of establishing a circle around the evader causes too large gaps between the agents through which the evader can escape.

\subsubsection{Local Observability}
The local observability case is studied with 20 agents and a communication cut-off distance of $d_c=40$.
Additionally, we introduce an observation radius $d_o=20$ within which the pursuers can observe the distance and bearing to the evader.
We reuse the \emph{basic} and \emph{extended} set from last section and modify the \emph{comm} set to include the shortest path information of other agents in the neighborhood of agent $i$ to the evader.
This way, each agent $i$ can compute a shortest path to the evader over a graph of connected agents, such that the path $P = (v^1, v^2, \dots, v^M)$ minimizes the sum $d_{\text{min}}^{i,e} =  \sum_{m=1}^{M-1}d^{m, m+1}$ where $v^1$ represents agent $i$ and $v^M$ is the evader.
The observation sets are given as:
\begin{align*}
	\emph{Basic}: \quad o^{i, j} &= \{d^{i, j},~ \phi^{i, j}\} & o^i_\text{loc} &= \{d^i_\text{wall},~ \phi^i_\text{wall},~ d^{i, e},~ \phi^{i, e}\} \\
	\emph{Extended}: \quad o^{i, j} &= \{d^{i, j},~ \phi^{i, j},~ \theta^{i,j} \} & o^i_\text{loc} &= \{d^i_\text{wall},~ \phi^i_\text{wall},~ d^{i, e},~ \phi^{i, e}\} \\
	\emph{Comm}: \quad o^{i, j} &= \{d^{i, j},~ \phi^{i, j},~ \theta^{i,j}, ~ d_{\text{min}}^{j,e} \} &  o^i_\text{loc} &= \{d^i_\text{wall},~ \phi^i_\text{wall},~ d^{i, e},~ \phi^{i, e},~ d_{\text{min}}^{i,e}\}.
\end{align*}
Note that in this case the distance and bearing to an evader are only available if $d^{i,e} \leq d_o$. Furthermore, the correct shortest path is only available if an agent and the evader are in the same sub-graph, otherwise, a pre-defined value is fed into the policy.

\begin{figure}
	\centering
	\includegraphics[width=0.5\columnwidth]{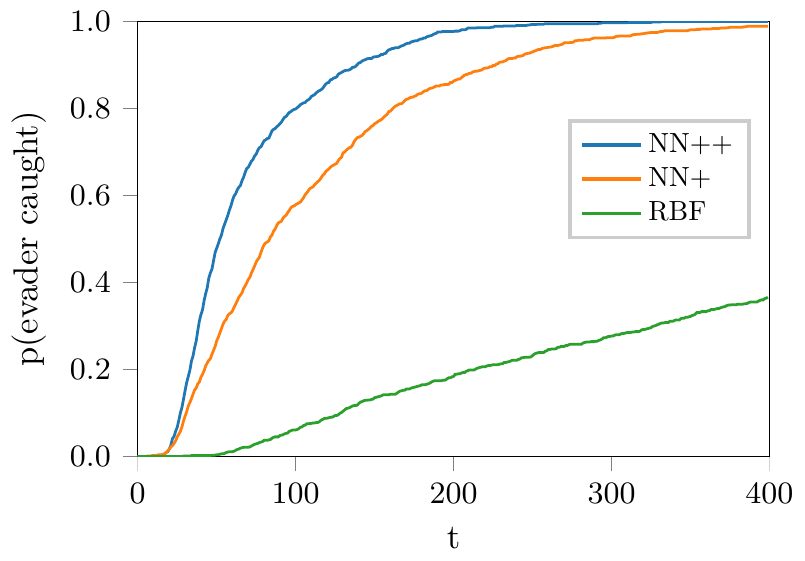}
	\caption{Performance comparison of the best policies in the pursuit evasion task with local observability. The curves show the probability that the evader is caught after $t$ time steps. All policies are learned and executed by 20 agents. Results are averaged over 1000 episodes with identical starting conditions.
	}
	\label{fig:pe_perf_comp_dig}
\end{figure}

Again, we limit the comparison for the local observability case to the more promising methods of neural network and RBF mean embeddings.
The results in Figure \ref{fig:pe_20a_40comm_top5_median} show that the performance gain of the neural network mean embeddings is even more noticeable than in the global observability case, with a clear advantage in the presence of the local communication protocols.
The inspection of the termination probabilities in Figure \ref{fig:pe_perf_comp_dig} confirms that the neural network mean embedding results in a significantly improved policy.

\subsection{Pursuit Evasion with Multiple Evaders}

Lastly, we study a pursuit evasion scenario with multiple evaders,  i.e., 
we assume that agent $i$ receives observation samples $\{o^{i,e}\}$ from several evaders, which are processed using a second mean embedding to account for the variable set size.
Where in the previous experiment the agents had precise information about the evader in terms of distance and bearing, they now have to extract this information from the respective embedding.
An additional level of difficulty results from the fact that the reward function no longer provides any guidance in terms of the distances to the evaders since it only counts the number of evaders caught in each time step (see Appendix \ref{appendix:rew_pe_multi} for details).

We study a scenario with 50 pursuers and 5 evaders using the global observability setup in Section~\ref{sec:evasionGlobalObs}, except that we respawn caught evaders to a new random location instead of terminating the episode.
The observation sets, containing the same type of information but arranged according to the inputs of the neural networks, are designed as follows:
\begin{align*}
\emph{Basic}: \quad o^{i, j} &= \{d^{i, j},~ \phi^{i, j}\} & o^{i, e} &= \{d^{i, e},~ \phi^{i, e}\} & o^i_\text{loc} &= \{d^i_\text{wall},~ \phi^i_\text{wall}\} \\
\emph{Extended}: \quad o^{i, j} &= \{d^{i, j},~ \phi^{i, j},~ \theta^{i,j} \} & o^{i, e} &= \{d^{i, e},~ \phi^{i, e}\} & o^i_\text{loc} &= \{d^i_\text{wall},~ \phi^i_\text{wall}\}.
\end{align*}

\begin{figure}
	\centering
	\captionsetup[subfigure]{margin={0.1cm,1cm}}
	\subcaptionbox{Evaluation of different observation models and embedding techniques. 
		\label{fig:pe_50a_top5_median}}{
		\includegraphics[scale=1]{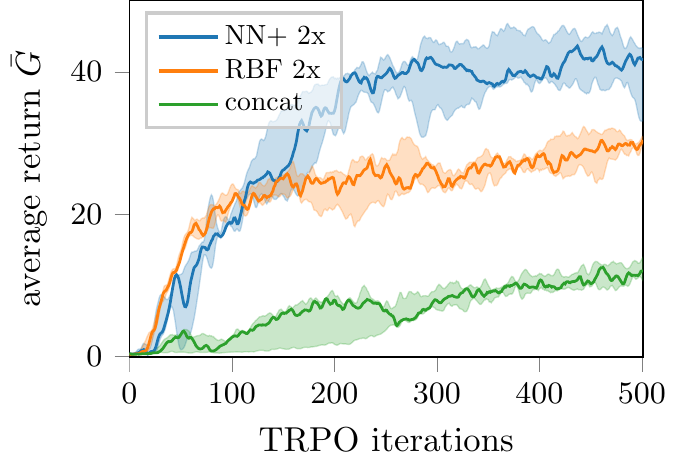}
	}
	\subcaptionbox{Comparison of the proposed mean embedding policy to policies with statistics of the observations as input.
		\label{fig:pe_50a_me_vs_stat_top5_median}}{
		\includegraphics[scale=1]{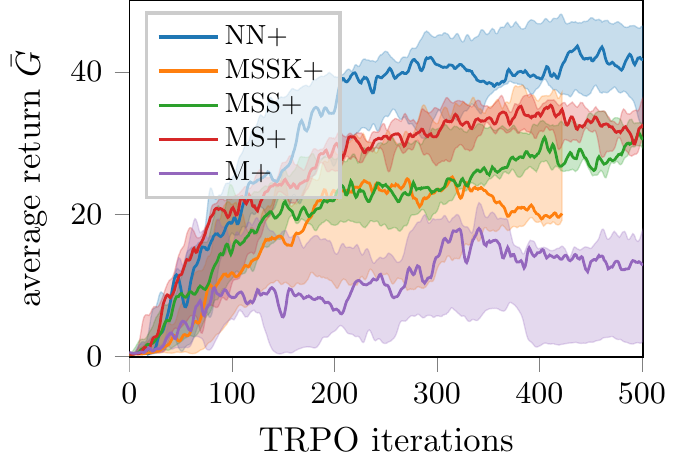}
	}	
	\caption{Learning curves for 50 agent pursuit evasion with 5 evaders. The curves show the median of the average return $\bar{G}$ based on the top five trials. \textbf{Legend:} NN+~2x: two neural network mean embeddings of the \emph{extended} set, RBF~2x: two radial basis function mean embeddings of the \emph{basic} set, concat: simple concatenation of \emph{extended} set. MSSK+, MSS+, MS+ and M+: Combinations of mean, standard deviation, skew and kurtosis of the features in the \emph{extended} set.}
\end{figure}

Figure \ref{fig:pe_50a_top5_median} shows the learning curves for policies with neural network and RBF mean embeddings and for the concatenation approach.
The return directly relates to the number of evaders caught during an episode.
Again, the neural network mean embedding performs significantly better than the RBF embedding.
The curves clearly indicate the positive influence of the additional information present in the \emph{extended} set.
With this amount of agents, the dimensionality of the concatenation has increased to a point where learning is no longer feasible.

\subsection{Evaluation of Pooling Functions}
\begin{figure}
	\centering
	\captionsetup[subfigure]{margin={0.1cm,1cm}}
	\subcaptionbox{Rendezvous with 20 agents and global observability.
		\label{fig:lc_rend_emb_comp}}{
		\includegraphics[scale=1]{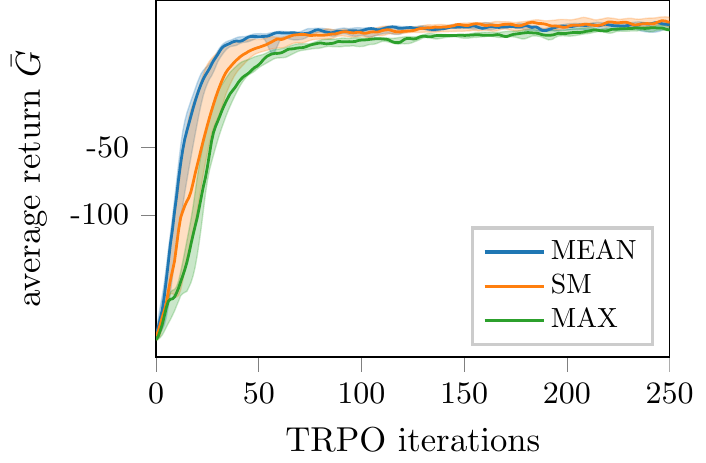}
	}
	\subcaptionbox{Pursuit evasion with 10 pursuers and global observability.
		\label{fig:lc_pe_emb_comp}}{
		\includegraphics[scale=1]{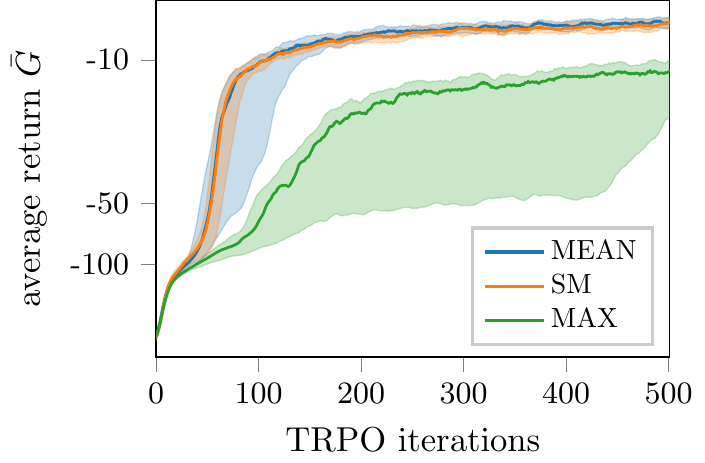}
	}	
	\caption{Learning curves of different embedding and pooling architectures based on the \emph{extended} set. The curves show the median of the average return $\bar{G}$ based on the top five trials on a log scale. \textbf{Legend:} MEAN: neural network mean embedding, SM: softmax feature pooling, MAX: max feature pooling.\label{fig:lc_emb_comp}}
\end{figure}

Figure \ref{fig:lc_emb_comp} shows learning curves of policies based on mean embeddings, softmax pooling, and max-pooling (as described in Section \ref{sec:repr_techs}) of features of the \emph{extended} set for the rendezvous and pursuit evasion task with global observability.

In the rendezvous task (Figure \ref{fig:lc_rend_emb_comp}), all pooling techniques eventually manage to find a good solution.
Policies using neural network mean embedding, however, on average converge more quickly while policies using max-pooling show slightly worse performance. Given its reduced computational complexity compared to the softmax-pooling, the mean embedding provides the most effective approach among all proposed architectures.

When examining the results of the pursuit evasion task (Figure \ref{fig:lc_pe_emb_comp}), we find that the algorithm produces two distinct solutions.
A sub-optimal one, which is only able to circle the evader but is unable to catch it (a catch is realized if the distance of the closest pursuer to the evader is below a certain threshold), and a solution which additionally catches the evader after a short period of time.
Therefore, we not only report the performance of the top 5 trials out of 16, but also provide the number of times the algorithm was able to discover the better of the two solution (Table \ref{tab:pe_emb_comp}).
Once that the algorithm finds a good solution, the mean embedding and softmax solutions perform comparably well but the max-pooling approach shows a significantly worse performance.
More importantly, however, the algorithm was able to find the good solution more often using the mean embedding than using the other pooling approaches.

\begin{table}
	\centering
	\begin{tabular}{c|c|c}
		mean & sm & max \\
		\hline
		10/16 & 6/16  & 4/16 
	\end{tabular}
	
	\caption{Number of times the algorithm discovered policies that led to a successful catch.\label{tab:pe_emb_comp}
	}
\end{table}

\subsection{Comparison to Moment-Based Representations}
Finally, we compare the mean embedding of neural network features to a representation using statistics of the input. 
Figure \ref{fig:pe_50a_me_vs_stat_top5_median} shows an evaluation on the pursuit evasion task with 50 agents and 5 evaders.
Here, we use combinations of the empirical mean, standard deviation, skew and kurtosis of each feature of the \emph{extended} set as the input to a policy function.
The plot reveals that neural network mean embeddings can capture more relevant information about the characteristics of the distribution of agents than simple statistics of the elements in the \emph{extended} set. Similar results were obtained for the other tasks although the differences in performance were less pronounced.

\subsection{Computational Complexity}
Unlike in classical optimization-based control, where the controller is derived from an assumed dynamics model, model-free reinforcement learning methods like TRPO 
find their control policies through interaction with the environment, without requiring explicit knowledge of the underlying system dynamics.
While this comes at the cost of an additional exploration phase, learning-based approaches typically offer an increased flexibility in that the same control architecture can adapt to different tasks and environments, without being affected by potential model mismatches.
More importantly, considering the final learned policy from a computational perspective, the synthesis of the control signal  involves no additional conceptual steps compared to an optimization-based approach.

While a typical experiment with 20 agents in our setup takes between four and six hours of training on a machine with ten cores (sampling trajectories in parallel), a forward pass through the trained neural network to compute the instantaneous control signal takes only about \SI{1}{\milli\second}, which enables an execution in real time. Furthermore, all control strategies learned through our framework are decentralized, which allows an arbitrary system size scaling in a real swarm network, where the required computations are naturally distributed over all agents.
When learning new policies, the memory requirements scale $\mathcal{O}(N(N - 1))$ with the number of agents (assuming global observability) since we need to store the local views of all agents. However, decentralized execution after the policy is learned scales linearly in $N$ per agent. An incremental online computation of the mean can be chosen if memory restrictions exist \citep{finch2009incremental}.

For comparison, the complexity of calculating Voronoi regions for the pursuit evasion policy scales $\mathcal{O}(N\log{}N)$ with the number of agents \citep{aurenhammer1991voronoi}.
Concerning the system sizes considered in our experiments, the resulting computation time of both policy types is in the same order of magnitude during task execution.

\section{Conclusion}
\label{sec:conclusion}
In this paper, we proposed the use of mean feature embeddings as state representations to overcome two major problems in deep reinforcement learning for swarms: the high and possibly changing dimensionality of information perceived by each agent.
We introduced three different approaches to realize such embeddings\Emdash two manually designed approaches based on histograms\textslash radial basis functions and an end-to-end learned neural network feature representation.
We evaluated the approaches on different variations of the rendezvous and pursuit evasion problem and compared their performances to that of a naive feature concatenation method and classical approaches found in the literature.
Our evaluation revealed that learning embeddings end-to-end using neural network features scales well with increasing agent numbers, leads to better performing policies, and often results in faster convergence compared to all other approaches.
As expected, the naive concatenation approach fails for larger system sizes.

\acks{
The research leading to these results has received funding from EPSRC under grant agreement EP/R02572X/1 (National Center for Nuclear Robotics). Calculations for this research were conducted on the Lichtenberg high performance computer of the TU Darmstadt.
}

\appendix

\section{Agent Kinematics}
\label{sec:app_kinematics}
In the single integrator case, the state of an agent is given by $s^i = [x^i, y^i, \phi^i] \in \mathcal{S} = \{[x, y, \phi] \in \mathbb{R}^3 : 0 \leq x \leq x_{\text{max}},\ 0 \leq y \leq y_{\text{max}}, \ 0 \leq \phi < 2\pi \}$, and the linear velocity $v$ and angular velocity $\omega$ can be directly controlled by the agent.
The kinematic model is given by
\begin{align*}
\dot{x} &= v \cos \phi \\
\dot{y} &= v \sin \phi \\
\dot{\phi} &= \omega.
\end{align*}
In the double integrator case, the state is given by $s^i = [x^i, y^i, \phi^i, v^i, \omega^i] \in \mathcal{S} = \{[x, y, \phi, v, \omega] \in \mathbb{R}^5 : 0 \leq x \leq x_{\text{max}},\ 0 \leq y \leq y_{\text{max}}, \ 0 \leq \phi < 2\pi, \lvert v \rvert \leq v_{\text{max}}, \lvert \omega \rvert \leq \omega_{\text{max}} \}$ and the agent can only indirectly change its velocity by acceleration.
With the control inputs $a_v$ and $a_\omega$, the model is then given by
\begin{align*}
\dot{v} &= a_v \\
\dot{\omega} &= a_\omega \\
\dot{x} &= v \cos \phi \\
\dot{y} &= v \sin \phi \\
\dot{\phi} &= \omega.
\end{align*}
For the experiments, we use finite differences to model the system in discrete time.

\section{Observation Model}
\label{sec:app_obs_model}
Irrespective of the task, an agent $i$ can sense the following properties about other agents $j \in \mathcal{N}(i)$ within its neighborhood:
\begin{align*}
d^{i, j} &  & \text{distance to neighboring agents} \\
\phi^{i, j} &= \arctan(\frac{y^j - y^i}{x^j - x^i}) - \phi^i & \text{bearing to neighboring agents} \\
\theta^{i, j} &= \arctan(\frac{y^i - y^j}{x^i - x^j}) - \phi^j & \text{relative orientation} \\
\Delta v^{i,j} &= v^i \left[\cos \phi^i, \sin \phi^i \right] - v^j \left[\cos \phi^j, \sin \phi^j \right] & \text{relative velocity}
\end{align*}
Furthermore, each agent has access to the following local properties:
\begin{align*}
d_{\text{wall}}^i &= \min(x^i - x_{\text{min}},~ y^i - y_{\text{min}},~  x_{\text{max}} - x^i, y_{\text{max}} - y^i) & \text{distance to closest wall} \\
\phi_{\text{wall}}^i &= \varphi_{\text{wall}}^i - \phi^i & \text{orientation to closest wall} \\
v^i, \omega^i & & \text{own velocity}
\end{align*}
where $\varphi^i_{\text{wall}}$ denotes the absolute bearing of agent $i$ to the closest wall segment.

\section{Task Specific Communication Protocols}
\label{sec:app_task_obs_model}
In the rendezvous task, agent $i$ additionally can sense information about neighborhood sizes:
\begin{align*}
\lvert \mathcal{N}(i) \rvert & & \text{own neighborhood size} \\
\lvert \mathcal{N}(j) \rvert: j \in \mathcal{N}(i) & & \text{neighborhood size of neighbor $j$}
\end{align*}
In pursuit evasion, we additionally have one or multiple evaders with states $s^e = \left[x^e, y^e\right] \in \{[x, y] \in \mathbb{R}^2 : 0 \leq x \leq x_{\text{max}},\ 0 \leq y \leq y_{\text{max}}\}$.
Agents can sense the distance and bearing to an evader, given that the evader is within an observation distance $d_o$:
\begin{align*}
d^{i, e} &=\sqrt{(x^i -x^e)^2 + (y^i - y^e)^2} \quad \text{if} ~ d^{i, e} \leq d_o & \text{distance to evader} \\
\phi^{i, e} &= \arctan(\frac{y^e - y^i}{x^e - x^i}) - \phi^i \quad \text{if} ~ d^{i, e} \leq d_o & \text{bearing to evader} 	
\end{align*}
Furthermore, we assume that each agent $i$ can compute a shortest path to the evader over a graph of connected agents, such that the path $P = (v^1, v^2 \dots, v^M)$ minimizes the sum $\sum_{m=1}^{M-1}d^{m, m+1}$ where $v^1$ is agent $i$ and $v^M$ is the evader.

\section{Controller for Double Integrator Dynamics}
\label{sec:app_pd_cont}
We use a simple PD-controller to transform the consensus protocol with high-level direct state manipulation to the unicycle model with double integrator dynamics.
It is given by

\begin{align*}
a_v &= K_1 (v_d - v) \\
a_\omega &= K_2 (\phi_d - \phi) + D_2 (\omega_d - \omega) \\
v_d &= \lVert \dot{\bm{x}} \rVert \\
\phi_d &= \arctan(\frac{\dot{y}}{\dot{x}}) \\
\omega_d &= 0,
\end{align*}
where the parameters $K_1$, $K_2$ and $D_2$ are tuned manually to give good performance on the problem.

\section{Reward Functions}
\label{appendix:rewardFunctions}

\subsection{Rendezvous}
\label{appendix:rew_rend}
The reward function is defined in terms of the inter-agent distances $\{d^{i, j}\}$ as
\begin{align*}
R(\bm{s}, \bm{a}) &= \alpha \sum_{i=1}^{N} \sum_{j=i+1}^{N} \min(d^{i, j}, d_c) + \beta \lVert \bm{a} \rVert,
\end{align*}
where in the global observability case we set the cut-off distance $d_c= \max(x_\text{max}, y_\text{max})$ to the maximum possible inter-agent distance in the respective environment.
The factor $\alpha = - \left(\frac{N(N - 1)}{2} d_c \right)^{-1}$ serves as a reward normalization factor and $\beta = \num{-1e-3}$ controls how strongly high action outputs of the policy are penalized.

\subsection{Pursuit Evasion}
\label{appendix:rew_pe}
For the case of a single evader, the pursuit evasion objective may be expressed in terms of the distance to the closest pursuer.
More specifically, the reward function is given as
\begin{align*}
R(\bm{s}, \bm{a}) &= - \frac{1}{d_o} \min(d_{\text{min}}, d_o),
\end{align*}
where $d_{\text{min}} = \min(d^{1, e}, \dots, d^{N, e})$.
For the global observability case, we set $d_o$ to the maximum possible distance of $d^{i, e}$ .

\subsection{Pursuit Evasion with Multiple Evaders}
\label{appendix:rew_pe_multi}
In the case of multiple evaders, we use a sparser reward function that  counts how many evaders are caught per time step, with no additional guidance of inter-agent distances.
An evader $e$ is assumed to be caught if the closest pursuer's distance $d_{\text{min}, e} = \min(d^{1, e}, \dots, d^{N, e})$ is closer than a threshold distance $d_t = 3$.
The reward function is given by
\begin{align*}
R(\bm{s}, \bm{a}) &= \sum_{e=1}^{E} \bm{1}_{[0, d_t]}(d_{\text{min}, e}),
\end{align*}
where $E$ is the number of evaders and
\begin{align*}
\bm{1}_{[a,b]}(x) &= \begin{cases}
1 & \text{if \,} x \in [a, b]\\
0 & \text{else}
\end{cases}
\end{align*}
is the indicator function.

\section{Policy Architectures}
\label{app:policy_archs}
This section briefly summarizes the chosen policy architectures.
Illustrations can be found in Figure \ref{fig:policy_models}.

\subsection{Neural Network Embedding Policy}
\begin{figure}
	\centering
	\captionsetup[subfigure]{margin={0.5cm,0cm},justification=RaggedRight}
	\subcaptionbox{Activation functions evaluation. 
		\label{fig:act_fct_eval}}{
		\includegraphics[scale=1]{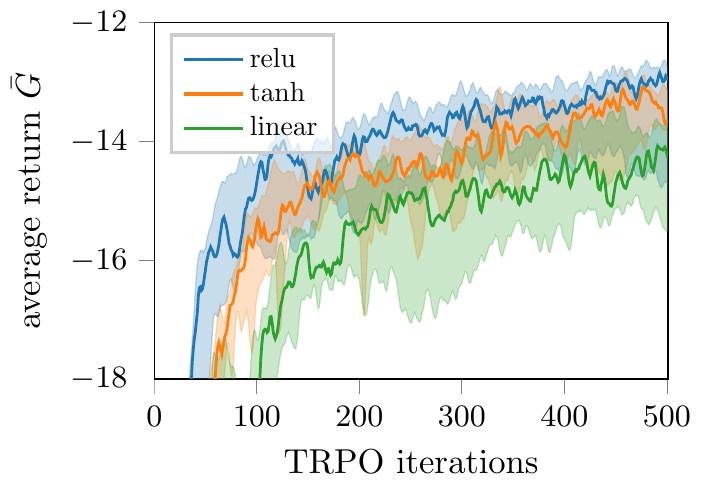}
	}
	\subcaptionbox{Layer size evaluation. 
		\label{fig:layer_eval}}{
		\includegraphics[scale=1]{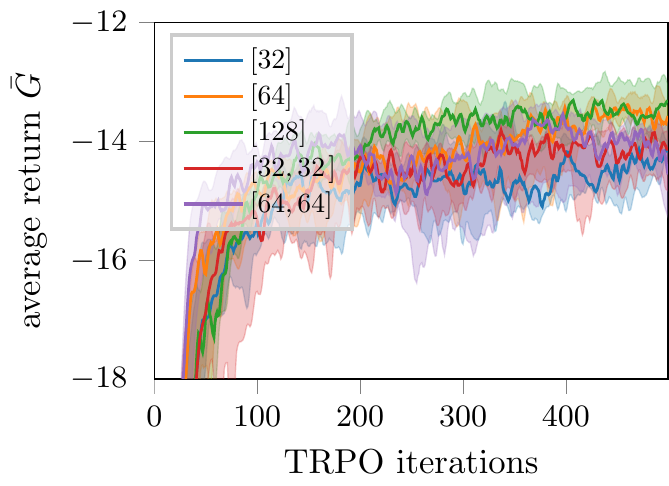}
	}	
	\caption{Learning curves for 20 agent rendezvous with (a) different activation functions for the mean embedding and (b) different layer numbers and sizes using a RELU activation function. The curves show the median of the average return $\bar{G}$ based on the top five trials.}\label{fig:arch_eval}
\end{figure}

We evaluated different layer sizes and activation functions on the rendezvous problem and show the results in Figure \ref{fig:arch_eval}. In all other experiments, the neural network mean feature embedding for agent $i$, given by
\begin{align*}
\phi^{\text{NN}}(O^i) = \frac{1}{\vert O^i \vert} \sum_{o^{i,j} \in O^i} \phi(o^{i, j}),
\end{align*}
is realized as the empirical mean of the outputs of a single layer feed-forward neural network,
\begin{align*}
\phi(o^{i, j}) &= h(W o^{i, j} + b),
\end{align*}
with 64 neurons and a RELU non-linearity $h$.

\subsection{Histogram Embedding Policy}
The histogram embedding is achieved with a two-dimensional histogram over the distance and bearing space to other agents.
We use eight evenly spaced bins for each feature, resulting in a 64 dimensional feature vector.

\subsection{RBF Embedding Policy}
The RBF embedding is given by a vector $\phi^\text{RBF}(O^i) = \left[\psi_1(O^i), \dots, \psi_{M^2}(O^i)\right]$ of $M^2$ contributions from $M=8$ radial basis functions whose center points are evenly distributed in the distance and bearing space.
With $o^{i,j} = [d^{i,j}, \phi^{i,j}]$, $\mu_m = [\mu_{d,m}, \mu_{\phi,m}]$, and $\sigma = [\sigma_{d}, \sigma_{\phi}]$ its components are given by
\begin{align*}
\psi_m(O^i) = \sum_{o^{i,j} \in O^i} \rho_m(o^{i,j}),
\end{align*}
where we choose
\begin{align*}
\rho_m(o^{i,j}) =
\exp\left(
-\frac{1}{2}\left[
\frac{(d^{i,j}-\mu_{d,m})^2}{\sigma_d^2} +
\frac{(\phi^{i,j}-\mu_{\phi,m})^2}{\sigma_\phi^2}
\right]
\right).
\end{align*}
The policy network structure used for both, the histogram and the RBF representations, is illustrated in Figure \ref{fig:policy_model2}.

\subsection{Concatenation Policy}
For the concatenation method, we first concatenate agent $i$'s neighborhood observations contained in the set $O^i$ and process them with one hidden layer of 64 neurons and a RELU non-linearity.
The resulting feature vector is then concatenated with the local properties $o^i_\text{loc}$ and fed into a second layer of same size.
Finally, the output of the second layer is mapped to the action.
The corresponding policy network structure can be seen in Figure \ref{fig:policy_model3}.

\FloatBarrier

\vskip 0.2in
\bibliography{me}   %

\end{document}